\documentclass[useAMS,usenatbib]{mn2e}
\usepackage{graphicx}
\usepackage{txfonts}
\usepackage{natbib}

\usepackage[english]{babel}
\usepackage{color}
\DeclareGraphicsExtensions{.pdf,.ps,.jpg}

\newcommand{\src}{SDSSJ143244.91+301435.3}
\newcommand{\sdss}{{\emph{SDSS}}}

\newcommand{\pedix}[2]{\ensuremath{#1_{\,\mbox{\scriptsize #2}}}}
\title[SDSSJ143244.91+301435.3: a link between RL NLS1 and CSS]{SDSSJ143244.91+301435.3: a 
link between radio-loud narrow-line Seyfert 1 galaxies and compact steep-spectrum radio sources?}
\author[Caccianiga et al.]{A. Caccianiga$^1$, S. Ant\'on$^{2,3}$, 
L. Ballo$^1$, D. Dallacasa$^{4,8}$, R. Della~Ceca$^1$, R. Fanali$^5$,
\newauthor  L. Foschini$^1$, T. Hamilton$^6$, A. Kraus$^7$,
T. Maccacaro$^1$, K.-H. Mack$^8$,  M.J. March\~a$^9$, 
\newauthor A. Paulino-Afonso$^2$, E. Sani$^{10}$, P. Severgnini$^1$ 
\vspace{0.2cm}
\\
   $^1$INAF - Osservatorio Astronomico di Brera, via Brera 28, 
 I-20121 Milan, Italy\\
  $^2$Faculdade de Ci\^encias da Universidade de Lisboa, Campo Grande, P-1749-016 Lisboa, Portugal\\
  $^3$Instituto de Astrof\'isica de Andaluc\'ia - CSIC, PO Box 3004, 18008 Granada, Spain\\ 
  $^4$Dipartimento di Astronomia, Universit\`a di Bologna, via Ranzani 1, 40127, Bologna, Italy\\
  $^5$Dipartimento di Fisica G. Occhialini, Universit\`a di Milano-Bicocca, Piazza della Scienza 3, 20126, Milano, Italy\\
  $^6$Department of Natural Sciences, Shawnee State University, 940 2nd Street, Portsmouth, OH 45662, USA\\
  $^7$Max-Planck-Institut f\"ur Radioastronomie, Auf dem H\"ugel 69, D-53121 Bonn, Germany\\
  $^8$INAF - Istituto di Radioastronomia, Via Gobetti 101, I-40129 Bologna, Italy\\
  $^9$32 Ordnance Hill, London NW8 6PU, UK\\
  $^{10}$INAF - Osservatorio Astrofisico di Arcetri, Largo E. Fermi 5, 50125, Firenze, Italy 
 }

   \date{}

\begin{document}

\label{firstpage}

\maketitle

\begin{abstract}

We present \src, a new case of radio-loud narrow line Seyfert~1 (RL NLS1) 
with a relatively high radio power
(P$_{1.4 GHz}$=2.1$\times$10$^{25}$ W Hz$^{-1}$) and large radio-loudness 
parameter (R$_{1.4}$=600$\pm$100). The radio source is compact
with a linear size below $\sim$1.4 kpc but, 
contrary to most of the RL NLS1 discovered so far with such a high R$_{1.4}$, 
its radio spectrum is very steep ($\alpha$=0.93, 
S$_{\nu}\propto\nu^{-\alpha}$) and not supporting a ``blazar-like'' nature. 
Both the small mass of the central super-massive black-hole 
and the high accretion rate relative
to the Eddington limit  estimated for this
object  (3.2$\times$10$^{7}$ M$_{\sun}$ and 0.27, respectively,
with a formal error of $\sim$0.4 dex on both quantities) are typical of 
the class of NLS1. 
Through a modelling of the spectral energy distribution of the source 
we have found that the galaxy hosting \src\ is undergoing a quite intense star-formation  
(SFR=50 M$_{\sun}$ y$^{-1}$) which, however, is expected to 
contribute only marginally ($\sim$1 per cent) to the observed radio emission. 
The radio properties of \src\ are 
remarkably similar to those of compact steep spectrum (CSS) 
radio sources, a class of AGN mostly composed by young radio galaxies. 
This may suggest a direct link between these two classes of AGN, 
with the CSS sources possibly representing the misaligned version (the 
so-called ``parent population'') of RL NLS1 showing blazar 
characteristics.

\end{abstract}

\begin{keywords}
galaxies: active -  galaxies: nuclei - quasars: individual: SDSSJ143244.91+301435.3
\end{keywords}

   \maketitle


\section{Introduction}
Narrow-line Seyfert 1 (NLS1) galaxies have been identified as a peculiar 
sub-class of AGN on the basis of their relatively narrow 
($\leq$2000 km s$^{-1}$) Balmer lines, 
relatively weak [OIII]$\lambda$5007\AA\ emission compared to the 
Balmer lines ([OIII]$\lambda$5007\AA/H$\beta<$3) and strong optical FeII 
(\citealt{Osterbrock1985, Goodrich1989, Pogge2000, Pogge2011, 
Veron-Cetty2001}). The presence of strong iron emission and the low 
[OIII]$\lambda$5007\AA/H$\beta$ flux ratio suggest that NLS1 are not obscured 
objects like type~2 AGN, where the observed narrow emission lines are produced in
regions (the Narrow Line Region, NLR) that are located far from the nucleus.
In NLS1, instead, the observed Balmer lines are likely produced in the 
Broad Line Region  (BLR) clouds that are located close 
($<$0.1 pc) the accreting super-massive black-hole (SMBH). 
If the BLR traces the potential well of the central 
SMBH, a low velocity dispersion is the result of a 
combination of a relatively small SMBH mass with  a high accretion rate, close
to the Eddington limit (L$_{Edd}$). Indeed, NLS1 are usually characterized 
by SMBH masses between 10$^{5}$ and 10$^{8}$ M$\sun$  
and accretion rates larger than $\sim$0.1L$_{Edd}$ 
(e.g. \citealt{Marziani2001, Boroson2002, Greene2004, Collin2004,  
Grupe2004a, Botte2004, Grupe2004, Zhou2006, Whalen2006,
Komossa2006, Yuan2008} but see also \citealt{Decarli2008, 
Marconi2008, Calderone2013}). 

An interesting characteristic of NLS1 is their preference of being 
radio-quiet (RQ) AGN. To date, only $\sim$50 radio-loud (RL) 
NLS1 have been discovered 
(\citealt{Foschini2011}). \citet{Komossa2006} has computed the 
fraction of radio-loud objects within the NLS1 population 
and compared it to the 
fraction of radio-loud sources within the population of broad line 
Seyfert~1 (BLS1) belonging to the same catalogue, and they found a significant 
difference ($\sim$7 per cent of radio-loud AGN, among NLS1, and 
20 per cent among BLS1). On a more general basis, 
the fraction of radio-loud AGN seems to depend both on the mass of the 
central SMBH (e.g. \citealt{Chiaberge2011}) and on the accretion rate 
relative to the Eddington limit: large SMBH masses combined to low accretion
rates are believed to give the highest probability of producing AGN with
large values of radio-loudness  while 
AGN characterized by high accretion rates and small SMBH masses, like
NLS1, seem to be less effective in producing radio-loud sources 
(e.g. \citealt{Lacy2001, Sikora2007}). 

A second characteristic of the few RL NLS1 discovered so far is 
that most of them present a compact radio emission (linear size below a 
few kpc, e.g.  \citealt{Doi2012} and references therein). 
Among the most radio-loud NLS1 there are many cases showing strict
similarities with the class of blazars (BL Lac objects 
and flat spectrum radio quasar, FSRQ): like blazars, these RL NLS1 
show a flat or inverted radio spectrum, high 
brightness temperatures ($T_b>$10$^7$ K, e.g. \citealt{Doi2013, Yuan2008}) and 
they are detected in gamma-rays by {\it Fermi}-LAT (\citealt{Abdo2009, 
Abdo2009a, Foschini2011}). 
Since blazars are usually believed to
be radio galaxies whose relativistic jets are pointing towards the observer 
(e.g. \citealt{Urry1995}), a possible conclusion is that also some of the 
RL NLS1 discovered so far are
``oriented'' and relativistically beamed sources. 
In this case we must expect a large number of mis-oriented and 
unbeamed sources, the so-called parent population, that in the standard 
beaming model is constituted by the class of lobe-dominated radio galaxies.
To date, however, only in six RL NLS1 an extended 
emission has been detected (\citealt{Whalen2006, Anton2008, Doi2012}) 
and only one RL NLS1  (SDSSJ120014.08--004638.7) is a 
lobe-dominated FR~II radio galaxy (\citealt{Doi2012}). 
It is not clear whether the lack of RL NLS1 in radio-galaxies 
is suggesting an intrinsic difference with respect to the other RL AGNs or 
if it is just a selection effect: if the BLR has a disk-like 
geometry, as suggested by some authors (e.g. \citealt{Pozonunez2013}, 
\citealt{Decarli2008}), 
then selecting AGN with the narrowest emission lines may preferentially 
lead to the selection of face-on systems and, 
therefore, with blazar-like radio morphology, 
in case of radio-loud AGNs. 
Alternatively, RL NLS1 are intrinsically different from 
more powerful RL AGN and lacking any extended emission. 
In this case, the mis-oriented population could appear as 
NLS1 with very weak (since unbeamed) radio cores i.e. sources classified 
as RQ NLS1 (\citealt{Foschini2013d}). Another interesting possibility is
that RL NLS1 are young or ``frustrated''  radio-galaxies that have not yet 
formed (or they are not able to form) radio lobes on large scales. This
possibility is suggested by the similarities found between the
class of RL NLS1 and that of the compact steep-spectrum (CSS) sources or
the gigahertz-peaked spectrum (GPS) sources (e.g. \citealt{Oshlack2001, 
Gallo2006, Komossa2006, Yuan2008}).

Clearly, a deeper investigation of the RL NLS1 and, in particular, of the
candidates for the parent population is mandatory to discriminate among
the various possibilties and to unveil the true nature of this class of AGN.
To date, only a few  RL NLS1 with non-blazar properties 
(e.g. with steep radio spectra)
have been discovered and studied. This number reduces to just 
few units if we require a value of the radio loudness parameter that 
unambiguously sets them in the class of radio-loud AGN (e.g. 
R$_{1.4 GHz}>$100, \citealt{Yuan2008, Komossa2006}).

In this paper, we present the discovery of a rare example of 
RL NLS1 (SDSSJ143244.91+301435.3, z=0.355) with a radio-loudness
parameter well in the radio-loud regime (R=160, computed at  
5~GHz, and R$_{1.4}\sim$600, computed at 1.4~GHz) and showing 
non-blazar properties. 
The study of this newly discovered
object may help in shedding light on the possible nature of the parent 
population
of this peculiar class of AGN.
In Section~2 we discuss the optical classification on the basis of the 
SDSS spectrum, 
while in Section~3 we study the properties of the host galaxy. 
In Section~4 we estimate the physical properties of the 
central SMBH, i.e. its mass and the accretion rate normalized to the 
Eddington limit. In Section~5 we analyse the radio data  
trying 
to disentangle the contribution from different sources present in the
field while in Section~6 we search for high-energy (gamma-ray and X-ray)
emission from \src\ by exploiting the existing  
public catalogues (ROSAT, XMM-{\it Newton}, {\it Fermi}-LAT). 
Discussion and conclusions are reported in Section~7.

Along the paper we assume a flat $\Lambda$CDM cosmology with H$_0$=71 km 
s$^{-1}$ Mpc$^{-1}$, $\Omega_{\Lambda}$=0.7 and $\Omega_{M}$=0.3. Errors are
given at 68 per cent confidence level.
   \begin{figure}
   \centering
    \includegraphics[width=6.5cm, angle=-90]{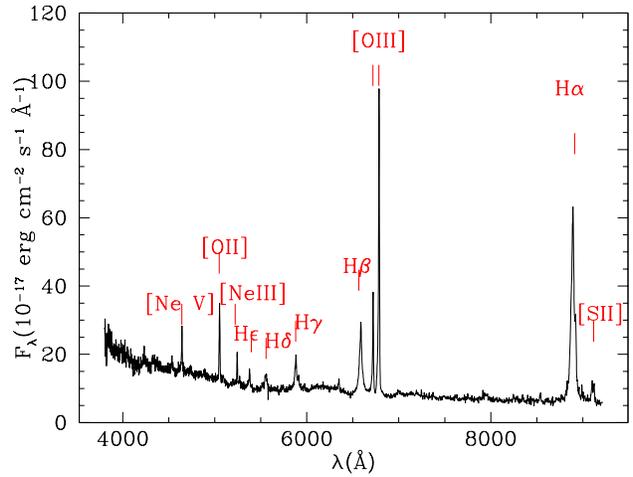}
   \caption{Optical (SDSS) spectrum in the observer's frame of 
SDSSJ143244.91+301435.3 (z=0.355). The strongest emission lines are labeled.
}
              \label{opt_spectrum}
    \end{figure}

   \begin{figure}
   \centering
    \includegraphics[width=6.5cm, angle=-90]{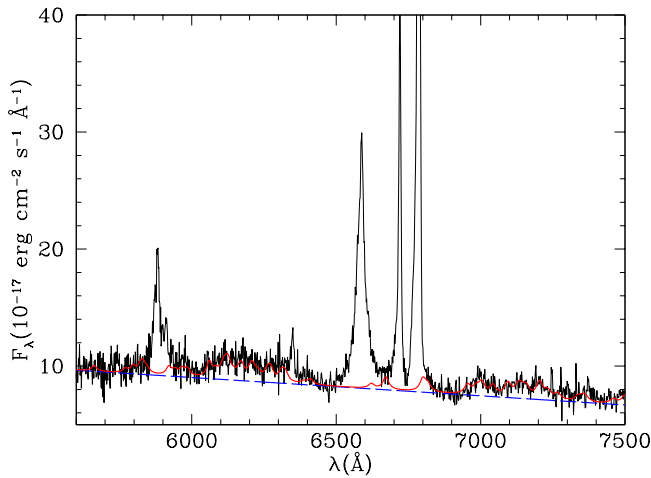}
   \caption{Result of the spectral analysis around the H$\beta$ region aimed at estimating the
intensity of the iron emission at 4570\AA. 
The dashed line (blue in the electronic version) is the fitted continuum while the solid line (red in the
electronic version) indicates the
continuum+iron emission based on the iron template presented in \citet{Veron-Cetty2004}.}
              \label{fe_fit}
    \end{figure}
   \begin{figure}
   \centering
    \includegraphics[width=6.5cm, angle=-90]{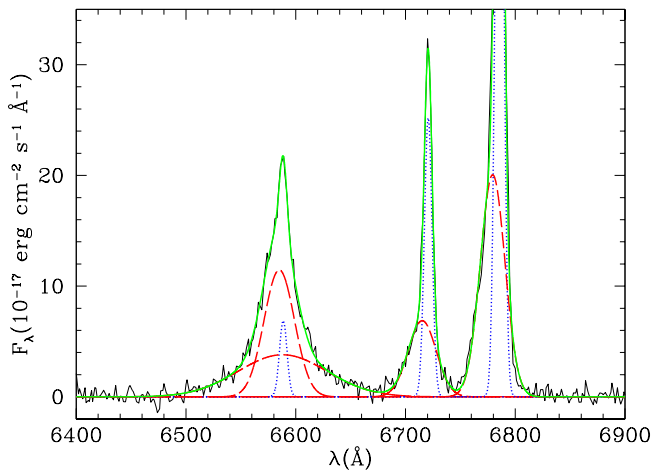}
   \caption{Iron and continuum subtracted spectrum around the 
H$\beta$ region. The results 
of the fitting procedure described in the text are also shown. Dotted lines are the narrow 
components (blue in the electronic version), dashed lines (red in the electronic version) 
represent the two broad components of H$\beta$ and the broad blue wings of the two [OIII] lines, 
the solid line (green in the electronic version) represents the total fit. 
}
              \label{iron_subtracted}
    \end{figure}

\section{Spectral analysis}

The source SDSSJ143244.91+301435.3 is spectroscopically 
classified as QSO at z=0.3548 in the Sloan Digital Sky Survey (SDSS, DR7). We have
analysed the optical spectrum (Fig.~\ref{opt_spectrum}) to derive a more detailed classification and to obtain 
some fundamental quantities, like the mass of the
SMBH (M$_{BH}$). In particular, we have
analysed the H$\beta$ spectral region which is critical to classify the source as NLS1 and also to derive a reliable SMBH mass using the single-epoch
(SE) method (\citealt{Vestergaard2006}).

Historically an AGN is classified as NLS1 when the H$\beta$ emission line is relatively 
narrow (FWHM$<$2000 km s$^{-1}$) compared to typical type~1 AGN. 
As type~1 AGN, instead, NLS1 have a low [OIII]$\lambda$5007\AA/H$\beta$ 
flux ratio ($<$3) that distinguishes them from absorbed (type~2) AGNs. 
Usually NLS1 present also strong
iron emission although this is not universally accepted as a defining property of the class. 
The strength of the iron emission at 4570\AA\, with  respect to the H$\beta$ line flux (broad plus 
narrow component), the so-called R4570 parameter, is typically larger than 0.4 
(\citealt{Veron-Cetty2001}). 

A simple fit with a Lorentian profile of the H$\beta$ yields a FWHM of 1370 km s$^{-1}$. 
However the probable presence of a narrow component of H$\beta$ 
may lead to an underestimate of the actual width of the broad component. An accurate determination of the
width of the broad component of H$\beta$ is required also to correctly derive 
M$_{BH}$.

In order to characterize the actual width of the broad component of H$\beta$ line and also
to estimate the intensity of the iron emission we analyse the spectral region around the H$\beta$ using
a multicomponent method similar to that described in \citet{Shen2011} and
\citet{Caccianiga2013}: 
first, we fit the spectrum in the 5600\AA-7500\AA\ range, avoiding the regions where the 
emission lines are present, using a combination of a power-law continuum 
and an iron pseudo-continuum template presented in \citet{Veron-Cetty2004} (see Fig.~\ref{fe_fit}).
Then, we subtract this template from the spectrum and fit the result
in the observed spectral range between 
6400-7000\AA\  using a model composed of several Gaussians: the two [OIII] 
lines are fitted with 4 Gaussians, 
two for the cores and two to fit the observed blue wings\footnote{The presence of blue wings in the [OIII] lines
is particularly important in NLS1 (see, e.g., \citealt{Veron-Cetty2001} and \citealt{Komossa2007})} 
(e.g. \citealt{Denney2009a, Shen2011}). 
The relative intensity of the two Gaussians 
describing the cores of the [OIII] emission lines is fixed to the theoretical value of 3:1.
The narrow component of H$\beta$ is described by one 
Gaussian whose width is forced to be the same  of the 
narrow [OIII]$\lambda$5007\AA\ line. 
For the broad component of H$\beta$ we initially use a single Gaussian which, however, does not reproduce
correctly the H$\beta$ profile. The fact that the broad
H$\beta$ in AGN is not well represented by a single 
Gaussian has been often pointed out (e.g. \citealt{Sulentic2000, Collin2006}) 
and it is particularly true for AGN belonging to the population~A i.e.
AGN with a broad component of the H$\beta$ with FWHM$<$4000 km s$^{-1}$. The profile is
usually fitted using a Lorentian function or several Gaussians.
We obtain a good fit to the data using two Gaussians. 
The lines positions are all independent, i.e. 
we have not fixed the relative distances. 
We have found that the relative positions of the cores of the narrow lines are
all consistent, within 6-7 km/s, with the expected values while,
as described below, we have found some 
significant offsets of the broad components with respect to the cores. 
The result of the fit is presented in Fig.~\ref{iron_subtracted}.

In this fit the broad H$\beta$ component (which is the
sum of two Gaussians) has a FWHM, corrected for the instrumental resolution,
of 1776 km s$^{-1}$, a value typical
for the NLS1 galaxies. The measured [OIII]$\lambda$5007\AA/H$\beta$ flux 
ratio (1.5) is in the range of values observed in 
unabsorbed AGN (type~1 AGN and NLS1).
From the spectral modelling we obtain also a value of the intensity of the
iron emission (R4570) equal to $\sim$0.45 
which is, again, in the typical range observed in NLS1 
(\citealt{Veron-Cetty2001}). 

The fitting procedure described above has clearly revealed the presence of 
broad wings on the blue side of
both [OIII] narrow lines (see Fig.~\ref{iron_subtracted}). 
The wings have a width (FWHM) of 1160-1340 km s$^{-1}$ and 
an offset relative to the line 
core\footnote{We note that, due to the intensity of the AGN emission, 
we do not observe any absorption line from the host galaxy in the optical 
spectrum. For this
reason we report the observed offsets relative to the cores of the narrow line
and not relative to an ``absolute'' reference frame like the one usually 
offered by the host galaxy lines} of 250-270~km s$^{-1}$. 
\citet{Veron-Cetty2001} found a blue wing in 13 out of 59 
NLS1 (22 per cent of the sample) with a range of widths 
between 525 and 1790 km s$^{-1}$ and offsets ranging from 90 to 570 km 
s$^{-1}$. 
It is believed that
these blue wings are originated from bi-conical structures that are outflowing
 from the nucleus with speeds 
of a few hundreds of km s$^{-1}$ while the narrow core of the [OIII] lines is 
expected to follow the
kinematics of the stars in the host galaxy (\citealt{Veron-Cetty2001}).
In AGN with a relativistic jet the structured shape of the [OIII] lines can 
also be
the result of the interaction of the relativistic jet with the interstellar 
medium (e.g. \citealt{Gelderman1994, Holt2008, Kim2013}).

Finally, we have analysed  
the  H$\alpha$ line. Similarly to what was done for the H$\beta$, 
we have modelled this line with 3 Gaussians, one for the
narrow component and two for the broad component. Two additional Gaussians
are used to model the [NII] lines at 6583\AA\ and 6548\AA, respectively. 
The relative intensity of these two lines has been fixed to the 
theoretical value of 2.96.
The width of all the narrow lines has been fixed to be equal to the line width
of [OIII] (adjusted for the spectral resolution). One of the two Gaussians 
describing the broad component of H$\alpha$ is 
found to be offset with respect to the narrow component ($\sim$10\AA\
corresponding to $\sim$330 km s$^{-1}$), although the significance of this 
offset is marginal. The FWHM of the broad component of H$\alpha$ is 
$\sim$1760 km s$^{-1}$ a value very similar to the width of the broad
component of the H$\beta$.

We conclude that SDSSJ143244.91+301435.3 can be confidently classified as 
NLS1 based on the H$\beta$ and H$\alpha$ line widths and on the basis of 
the [OIII]$\lambda$5007\AA/H$\beta$ flux ratio. Moreover, the source 
presents other peculiar properties, like the intensity of the iron emission 
and the presence of blue wings in the [OIII] 
emission lines, that are typical of this class of AGN.

\section{Properties of the host galaxy}
One important issue related to the RL NLS1 is about the galaxy type hosting 
these AGN. On the one hand, RQ NLS1 are usually 
associated with spiral hosts that often show intense star-forming activity 
(e.g. \citealt{Deo2006, Sani2010, Sani2012}) and, on 
the other hand, radio-loud objects (at least the most radio-loud ones) 
are usually hosted by early-type galaxies (e.g. \citealt{Sikora2007} 
and references therein). 

The origin of the dichotomy in the morphology of the host galaxy of RQ and RL 
AGN is not yet clear but there
are indications suggesting that it could be related to the 
different accretion history
of the two classes of galaxies that eventually affects the final value of the 
spin of the central SMBH: many small mass accretion events in disk 
galaxies and major mergers in ellipticals (\citealt{Volonteri2007, Sikora2009}). 
The latter process is expected to
produce more easily large values of spin in the central SMBH a condition that
is likely connected to a greater efficiency in producing relativistic jets 
(``spin paradigm'', {\citealt{Blandford1990}).

The peculiar combination of a radio-loud AGN with NLS1 
properties opens the question of what kind of galaxy is hosting these 
objects. So far, only in very few RL NLS1 the host galaxy has been studied
and characterized (\citealt{Zhou2007, Anton2008, Hamilton2012, Foschini2012}). 
These observations have revealed the 
existence of arms or circumnuclear rings (possibly the consequence of a recent
merger) and the presence of a circumnuclear starburst. 
These results seem to suggest that the galaxies 
hosting these RL NLS1 are more similar to those typically observed in the 
radio-quiet counterparts but it is still premature to
derive any firm conclusion. 

For the reasons explained above, it is 
interesting to study the properties of the host galaxy of \src.
We do it, both using the SDSS images and by analysing the 
Spectral Energy Distribution (SED) of the object. 

\subsection{Radial profile}

   \begin{figure}
   \centering
    \includegraphics[width=7cm]{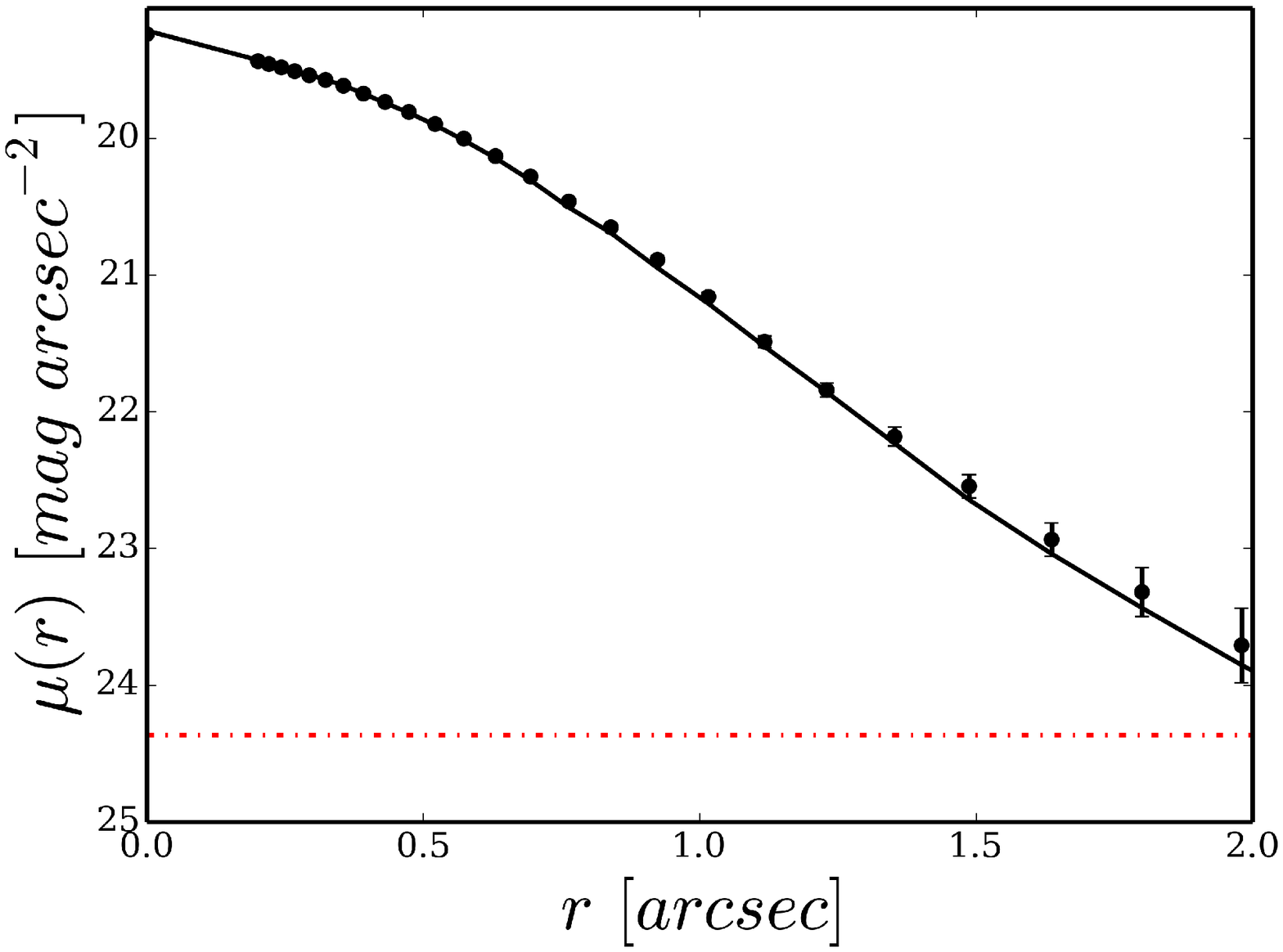}
    \includegraphics[width=7cm]{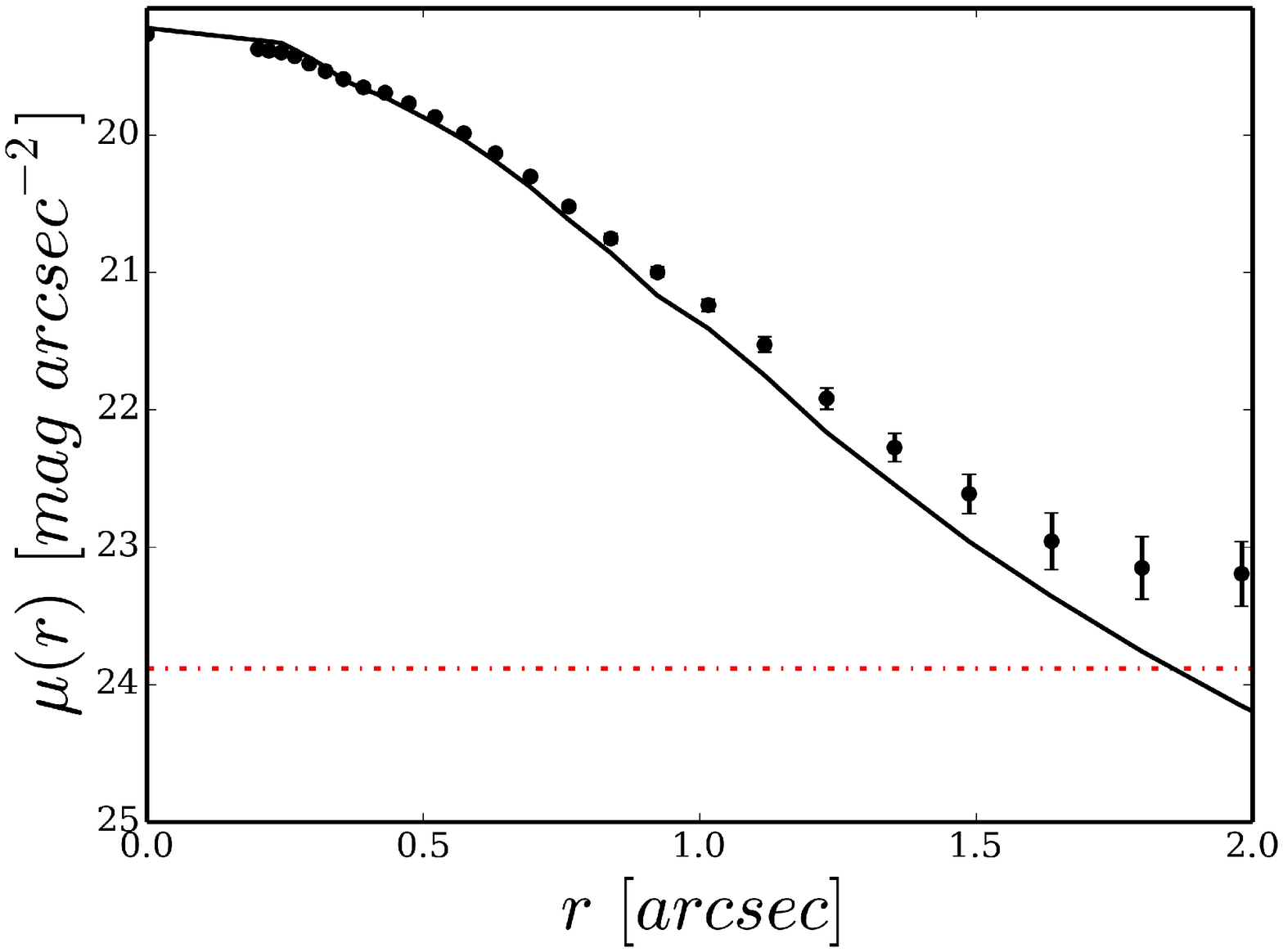}

   \caption{Radial profile in the SDSS $r$-band (top) and in the $i$-band (bottom) 
of \src. The PSF profile is indicated with the solid line while the red dashed dotted 
line represents the background.
}
              \label{gal_profile}
    \end{figure}


We have analysed the multi-band SDSS-DR9 images.
A 2-D modelling, using the GALFIT package (\citealt{Peng2002, Peng2010}), was performed 
in each band, where different Sersic profiles were tentatively fitted. 
The results show that the object is best  modelled by a PSF. The latter is reconstructed from 
the images provided by SDSS, on all five bands.
Similar results were found after testing different PSF models, namely by building our PSF from the 
field image 
using the IRAF daophot package and by stacking bright, non-saturated and isolated stars of the field. 
In Fig.~\ref{gal_profile} we present the 1-D profiles in $r$ and $i$-band. 
The surface brightness profiles were derived from the ellipse 
STSDAS routine and then converted to $\mathrm{mag\  arcsec^{-2}}$ via 

\begin{equation}
\mu = -2.5 Log\left(\frac{I}{s^2 t_{exp}}\right) + m_{zpt}
\end{equation}
where $I$ is the output profile from ellipse, s=0.396 arcsec/pixel is the plate scale of the SDSS, $t_{exp}$ is the exposure time and $m_{zpt}$ is the magnitude zero-point all extracted from the image header. 
The error bars of the profile include two distinct terms. One is the propagated error from the derived flux profile 
and the second term takes into account the background noise. 
The model profiles match the observed profiles closely with the exception of the $i$-band profile where a greater deviation from the 
PSF profile might be indicative of an excess of emission from the host galaxy (see Fig.~\ref{gal_profile},
lower panel) 
at the wavelength of 7625\AA\ (corresponding to 5627\AA, rest-frame).
However, GALFIT cannot converge on any meaningful solution, thus, no structural parameters can be derived.

\subsection{Optical and IR SED modelling}

The analysis of the optical images of \src\ has not allowed us to derive any
information about the type of host galaxy but it suggested that in the reddest
part of the spectrum the galaxy could be becoming more and more important.
This means that a proper analysis of the Spectral Energy Distribution (SED) of \src,
including infrared (IR) photometric data, may reveal some useful hints about the type of host galaxy. 
\begin{figure*}
 \centering
\resizebox{0.48\hsize}{!}{\includegraphics{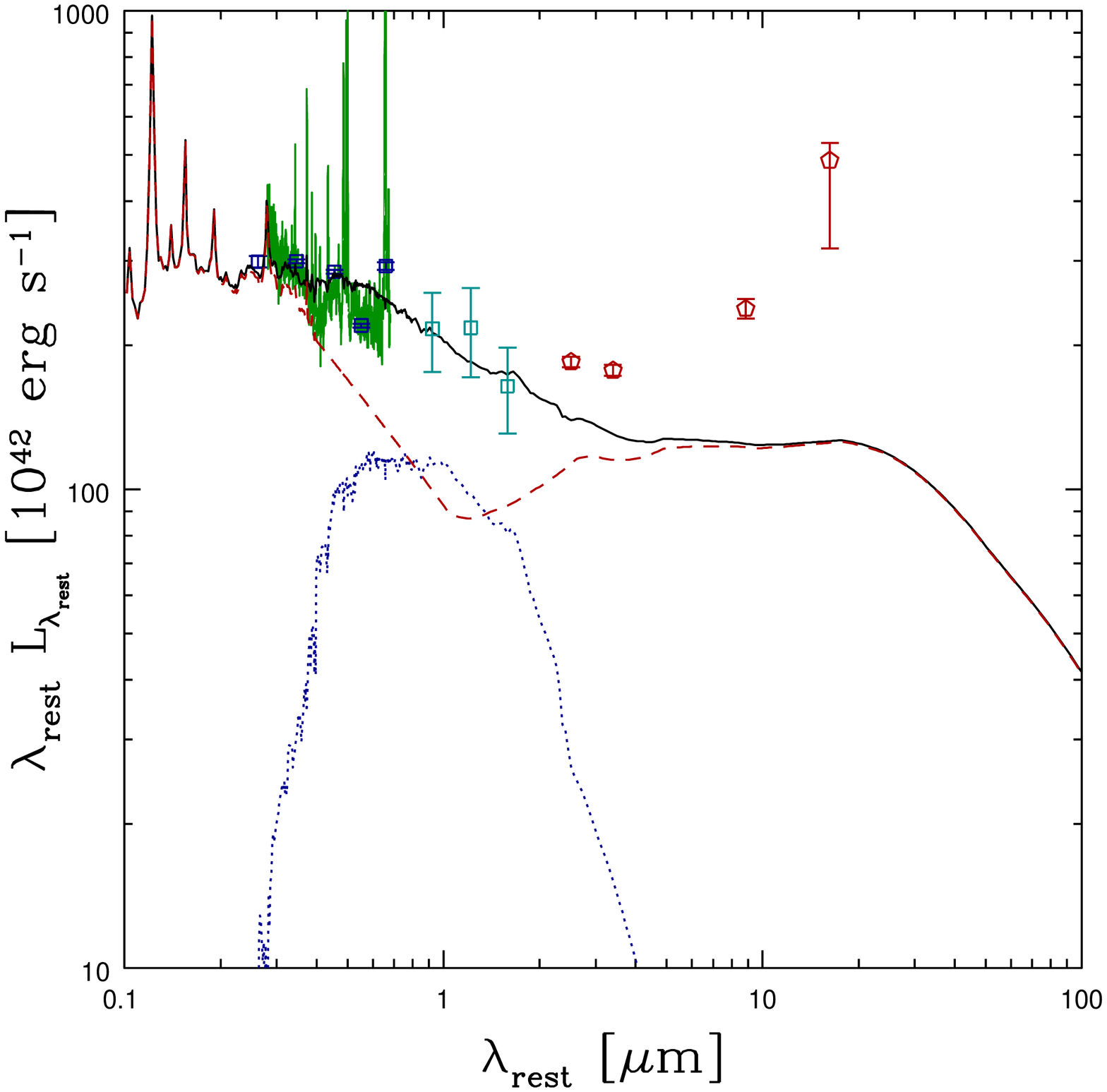}}
\resizebox{0.48\hsize}{!}{\includegraphics{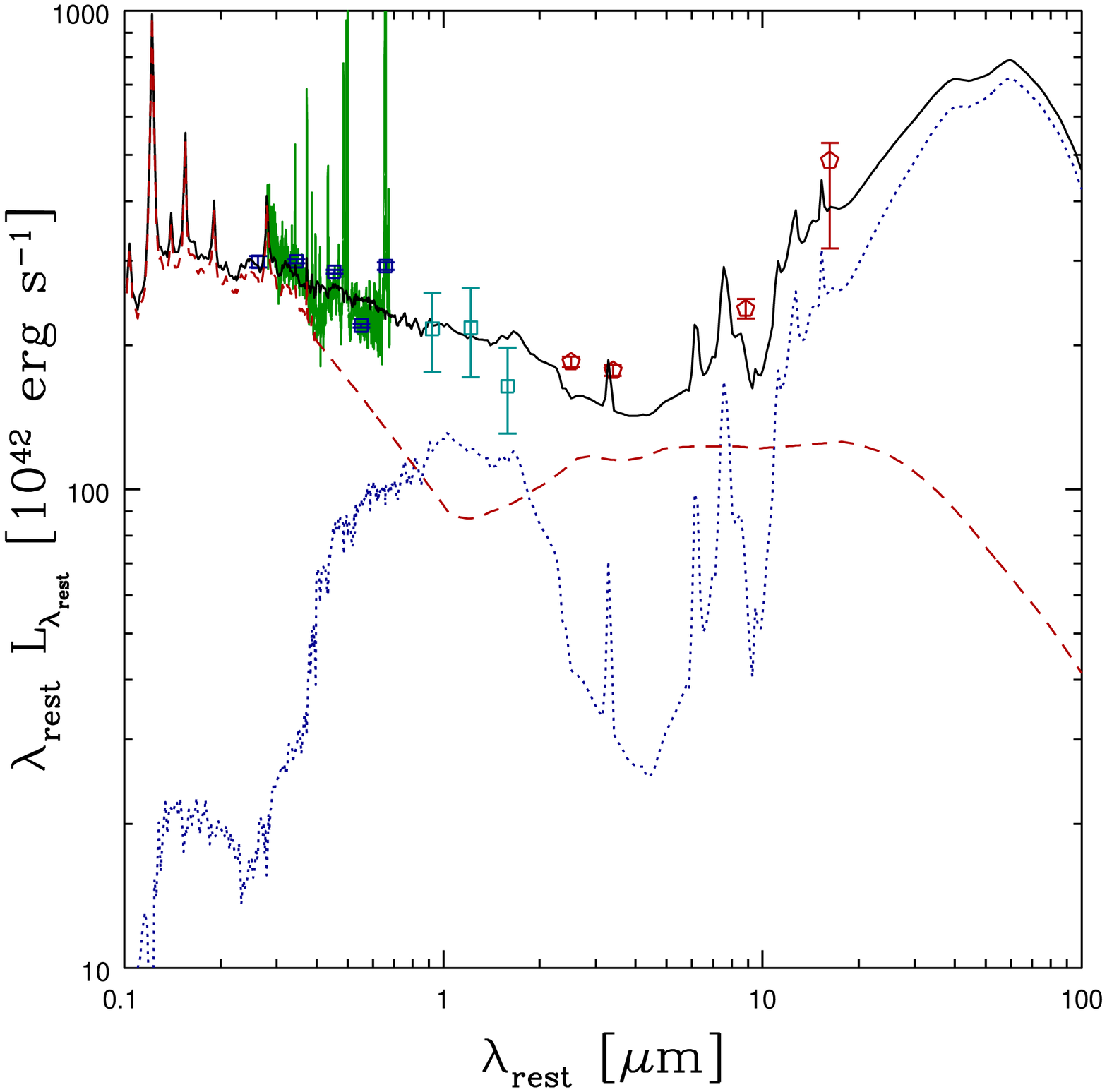}}
 \caption{Spectral Energy Distribution of \src, in the visible/IR range. 
Photometric points
are taken from SDSS (visible; blue points in the colour version), 2MASS 
(NIR; cyan points in the colour version) 
and WISE (MIR; red points in the colour version). The optical (SDSS) spectrum 
is also plotted (green line in the colour version). The solid line is the 
model that better reproduces the data, and it is composed of an AGN template 
(BQSO1 from \citealt{Polletta2007}, dashed red line) plus a galaxy template 
(dotted blue line): in the left panel an elliptical galaxy template is used 
while, in the right panel, we adopt a starburst galaxy template (M82). 
See text for details.}
 \label{fig:sed}%
\end{figure*}
We have thus built the SED of \src\ from optical to IR wavelengths 
using the photometric points from the available
catalogues. Besides the \sdss\ photometric data, magnitudes in the 
$J$, $H$, and \pedix{K}{s} 2MASS bands are available (see Tab.~\ref{tab_ir}).
At longer wavelength, IR information are provided by the Wide-field Infrared 
Survey Explorer (\citealt{Wright2010}), a 
NASA satellite that imaged the whole sky in four mid-IR photometric bands, 
centred at $3.4$, $4.6$, $12$ and $22\,\mu$m.
We refer to these bands as W1, W2, W3 and W4, respectively. 
A WISE object, at a distance of $0.17\arcsec$ from the optical position of 
\src, is detected in all four 
bands with high significance ($S/N$ of $38.9$, $37$, $24.7$, and $12.8$ 
in W1, W2, W3 and W4, respectively).
In Tab.~\ref{tab_ir} we report the profile-fit photometry as obtained from 
the WISE All-Sky source catalog; the 
corresponding flux densities have been computed by assuming the magnitude zero 
points of the Vega system corresponding
to a power-law spectrum ($f_{\nu} \propto \nu^{\alpha}$) with $\alpha=-1$. 
The differences in the computed flux densities expected using flux correction 
factors that correspond to $\alpha=1,0$ or
$-2$ (lower than $0.8$ per cent, $0.6$ per cent, $6$ per cent, and $0.7$ per 
cent in W1, W2, W3 and W4, 
respectively) have been added to the
catalogued flux errors. 
To account for the observed discrepancy between the red and blue calibrators 
used for the conversion from magnitudes to 
Jansky, an additional $10$ per cent uncertainty was added to the $12$ and 
$22\,\mu$m 
flux densities (\citealt{Wright2010}).

\begin{table*}
 \caption{IR data.}             
 \label{tab_ir}     
{
  \begin{tabular}{l r@{$\pm$}l r@{$\pm$}l r@{$\pm$}l l l r@{$\pm$}l r@{$\pm$}l r@{$\pm$}l r@{$\pm$}l}
   \hline\hline
    \multicolumn{7}{c}{2MASS} & & \multicolumn{9}{c}{WISE} \\
    \multicolumn{1}{c}{Name} & \multicolumn{2}{c}{$J$} & \multicolumn{2}{c}{$H$} & \multicolumn{
2}{c}{\pedix{K}{s}} &  & \multicolumn{1}{c}{Name} & \multicolumn{2}{c}{$3.4\,\mu\mbox{m}$} & 
\multicolumn{2}{c}{$4.6\,\mu\mbox{m}$} & \multicolumn{2}{c}{$12\,\mu\mbox{m}$} & \multicolumn{2}{c}
{$22\,\mu\mbox{m}$} \\
    \multicolumn{1}{c}{(1)} & \multicolumn{2}{c}{(2)} & \multicolumn{2}{c}{(3)} & 
\multicolumn{2}{c}{(4)} & & \multicolumn{1}{c}{(5)} & \multicolumn{2}{c}{(6)} & \multicolumn{2}{c}{(7)} & 
\multicolumn{2}{c}{(8)} & \multicolumn{2}{c}{(9)} 
\vspace{0.1cm} \\
\cline{1-7}\cline{9-17}
\hfill \break
  958046664 & $17.20$&$0.21$ & $16.42$&$0.23$ & $15.88$&$0.22$ & & J143244.92+301435.3 & $
14.46$&$0.03$ & $13.54$&$0.03$ & $10.32$&$0.04$ & $7.46$&$0.09$ \\ 
   \hline
 
  \end{tabular}
}
 {
\hfill \break
\footnotesize Col.~(1): 2MASS All-Sky Release PSC key.
 \footnotesize Col.~(2)-(4): 2MASS magnitudes, not corrected for Galactic extinction.
 \footnotesize Col.~(5): WISE source designation.
 \footnotesize Col.~(6)-(9): WISE magnitudes, not corrected for Galactic extinction.}
\hfill \break
\end{table*}

In order to deconvolve the galaxy and AGN contributions, we adopted a very 
simple phenomenological approach: 
The SED has been modelled by the sum of a galaxy and an AGN template 
chosen from the SWIRE template library of \citet{Polletta2007}.
We overplotted
the model on the SED and chose the combination of templates that better
reproduced the data.
The AGN templates have been derived by combining the \sdss\ quasar composite 
spectrum and rest-frame IR data of a sample
of optically-selected type~1 QSOs observed in the SWIRE program.
We considered three templates with the same optical spectrum but three 
different IR SEDs: a mean IR spectrum, obtained
from the average fluxes of all measurements (``QSO1''), a template with 
high IR/optical flux ratio template,
obtained from the highest $25$ per cent measurements per bin (``TQSO1''), 
and a 
low-IR emission SED obtained from the lowest
$25$ per cent measurements per bin (``BQSO1'').
The continuum and the broad line components of the AGN template were absorbed 
with the extinction curve taken from
\citet{Chiar2006}. The best representation of the AGN continuum 
is obtained using an A$_V$ of 0.1 mag. 
For the galaxy emission, we considered $16$ templates, covering the 
wavelength range between $1000\,$\AA\ and $1000\,\mu$m.
They include $3$ ellipticals, $7$ spirals (from early to late types, S0-Sdm), 
and $6$ starburst templates.
When overplotted on the top of the spectrum, the \sdss\ photometric points 
look fairly consistent (see
Fig.~\ref{fig:sed}, left panel), suggesting that the $3\arcsec$ fiber 
includes most of the source flux.
Therefore, we included also the spectrum in the SED modeling.
At short wavelength, the observed fluxes are clearly dominated by the emission 
from the active nucleus.
In order to correctly describe the optical spectrum and the \sdss\ photometric 
points without overestimating the emission at 
$3.4$, $4.6$, and $12\,\mu$m, we have to adopt the ``BQSO1'' template.
Starting from the $i$ and $z$ \sdss\ bands the observed SED requires 
an additional contribution that becomes particularly important in the 2MASS 
energy range. This contribution is probably due to the emission from the host 
galaxy that
is more and more important in the red part of the SDSS spectrum. This
is independently confirmed by the analysis of the radial profile of \src,
discussed in the previous section, where we have found that in the $i$-band
there is an indication of the presence of some extended emission
(contrary to what is observed at shorter wavelengths).
While any host template is able to reproduce the photometry observed up 
to $\sim 4\,\mu$m (rest frame), the W3 and
particularly the W4 data clearly require a galaxy with a significant
star-formation (SF) emission. The best representation of
the data (see Fig.~\ref{fig:sed}, right panel) is obtained using the 
template of M82. 

It is interesting to estimate the star-formation rate (SFR) from the 
intensity of
the IR emission of the host galaxy. From \citet{Kennicutt1998} 
we can infer the SFR from the integrated
8-1000$\,\mu$m luminosity (L$_{FIR}$):

\begin{equation}
\frac{SFR}{M_{\sun} y^{-1}} =4.5\times10^{-44} \frac{L_{FIR}}{erg s^{-1}}
\end{equation}

By integrating the M82 template, normalized as in Fig.~\ref{fig:sed}, we obtain a 
L$_{FIR}$=1.13$\times$10$^{45}$ erg s$^{-1}$ that implies a 
SFR of $\sim$50.8 $M_{\sun}$ y$^{-1}$.
Given the value of L$_{FIR}$ inferred from the SED, \src\ can be classified as Luminous 
Infrared Galaxy (LIRG, e.g. \citealt{Sanders1996}).

The presence of a star-forming host galaxy is in agreement with what is
usually found in RQ NLS1. In particular, \citet{Sani2010} has found that 
SF is usually higher in NLS1 galaxies than in BLS1 galaxies 
with the same nuclear luminosity. Using our SED modelling we have estimated
the SF vs AGN luminosity ratio as measured by the
parameter (\citealt{Sani2010}):

\begin{equation}
R_{6} = \frac{L(6.2, PAH)}{\nu L_{\nu} (6, AGN)} 
\end{equation}

where $L(6.2, PAH)$ is the luminosity of the 
polycyclic aromatic hydrocarbon (PAH) line at 6.2$\,\mu$m and
$\nu L_{\nu} (6, AGN)$  is the AGN luminosity at 6$\,\mu$m.
We obtain  $R_{6}$=0.02 which is in the typical range of
values observed in NLS1 galaxies and close to the median value measured 
for this class of AGN. On the contrary, BLS1 usually present
values of  $R_{6}$ below 0.01 (e.g. see Fig.~3 in \citealt{Sani2010}). 
We note that the value of R$_6$ reported here is just an estimate based on
the SED modelling and it is not  computed from an IR spectrum. 
In particular, the equivalent widths of the lines in the real spectrum 
could be different from those in the adopted template. However, the
good agreement between the estimated R$_6$ and the value expected in NLS1
can be considered as a support of the reliabilty of our SED
modelling.

\subsection{Independent estimate of the extinction}
The SED modelling described in the previous section has revealed a
low value of optical extinction (A$_V\sim$0.1 mag) in the AGN component. 
This value is relatively
well constrained since, with A$_V$ greater than
$\sim$0.3 mag, we cannot reproduce the SED in an acceptable way. 
At the same time,
the SED modelling requires the presence of a high level of 
SF (that is usually associated to high values of
extinction).
The two results are not in contradiction since SF 
in the host
of NLS1 is often observed in extended structures, like bars 
(e.g. \citealt{Deo2006}), that do not necessarily intercept the line of 
sight between the observer and the active nucleus.  Therefore, 
the AGN could be relatively free from reddening even if the host
galaxy is undergoing an intense SF.
In order to have an independent confirmation that the nuclear extinction is
actually low, we have used the spectral analysis presented in Section~2 
to estimate the A$_V$ from the Balmer decrement i.e. the 
ratio between H$\alpha$ and H$\beta$ fluxes. Since the narrow emission lines 
are expected to be contaminated by the SF emission\footnote{Using the
relation between SFR and H$\alpha$ luminosity in
star-forming galaxies (\citealt{Kennicutt1998}) we have estimated
that more than 80 per cent 
of the narrow component of the H$\alpha$ could be produced 
by the SF (assuming SFR=50~M$_{\sun}$ y$^{-1}$)}, we have used
the broad emission lines. 
It should be noted, however, that the determination of the amount of 
reddening using the broad emission lines is difficult because of 
the high uncertainty on the theoretical value of the intrinsic 
H$\alpha$/H$\beta$ flux ratio in the BLR. While this value is
quite well determined for the NLR of AGN
(F(H${\alpha}$)/F(H${\beta}$)=3.1, e.g. \citealt{Gaskell1984}), in the
BLR it is expected a value larger than 3.1, as discussed in 
\citet{Maiolino2001}. Therefore, the 
A$_V$ computed from the broad emission lines,  
assuming an intrinsic F(H${\alpha}$)/F(H${\beta}$)=3.1, 
should be considered
only as an upper limit to the actual value.
Using the broad components of the H$\alpha$ and H$\beta$ lines 
as derived from the spectral analysis presented in Section~2 and
corrected for the Galactic reddening, we have 
computed a Balmer decrement of 2.7$\pm$1. 
Considering the maximum value of this ratio at 1$\sigma$ (3.7), and 
assuming an intrinsic decrement of 3.1 we obtain an upper limit
on the A$_V$ of $\sim$0.5 mag. 
We thus confirm that the nuclear extinction is relatively
low and consistent with what has been found 
in the SED modelling.

As mentioned above, we expect that the SF contributes
significantly to the observed narrow emission lines. Therefore,  
the Balmer decrement computed using the narrow lines could give 
indications of the extinction in the regions where the SF occurs. 
The observed Balmer decrement using the narrow 
components (corrected for the Galactic reddening) 
is 4.1$\pm$1.8. The large uncertainty on this value is
due to the difficulty of measuring the narrow components of the lines, 
in particular of the H$\alpha$. Using this Balmer decrement we  estimate
A$_V\sim$0.9 mag (with a 1$\sigma$ uncertainty interval of 0 -- 2 
mag). 
A value of A$_V\sim$0.9 mag is consistent with what
is usually observed in starburst galaxies 
(e.g. \citealt{Kennicutt1998, Buat2002, 
Dominguez2013, Ibar2013}) and this confirms that a significant 
fraction of the observed narrow H$\alpha$ and H$\beta$ is probably produced 
in the star-forming regions of the host galaxy.

\section{Properties of the SMBH}

From the H$\beta$ width we can compute the mass of the SMBH
using the single-epoch relation presented in \citet{Vestergaard2006}:

\begin{equation}
Log\left(\frac{M_{BH}}{M_{\sun}}\right) = 6.91 + 2 Log \left(\frac{FWHM (H\beta)}{1000 km s^{-1}}\right) + 0
.50 Log \left(\frac{\lambda L_{5100\AA}}{10^{44} erg s^{-1}}\right)
\end{equation}

using the observed L$_{5100\AA}$ (2.17$\times$10$^{44}$ erg s$^{-1}$) and the
FWHM of the broad component of H$\beta$ (1776 km s$^{-1}$) 
we find M$_{BH}$=3.8$\times$10$^{7}$ M$_{\sun}$.
The statistical (1$\sigma$) uncertainty on the mass, as propagated 
from the errors on the two input variables (the line width 
and the 5100\AA\ luminosity) is 10\%. 
We note, however, that besides the statistical error there is an additional
source of uncertainty related to the single-epoch method and that is mainly
connected to the unknown geometry and orientation of the BLR  
(see discussion in \citealt{Vestergaard2006}). 
This uncertainty, estimated to be about 0.35-0.46 dex, i.e. a factor 2-3 
(\citealt{Park2012}), is much larger than the statistical uncertainty 
and it should be considered as a more reliable estimate
of the actual error on the computed mass. 
 
The observed luminosity at 5100\AA\ (rest-frame) is likely contaminated 
by the host galaxy and, therefore, the actual SMBH mass is expected to be lower. At the same
time, a correction for the extinction would increase the value of mass. 
Using the results from the SED modelling described in Section~3, we expect a
contribution of the host galaxy to the
luminosity at 5100\AA\ of $\sim$33 per cent and an extinction of A$_V\sim$0.1 mag.
If we correct the observed  
L$_{5100\AA}$ for both effects we obtain 
L$_{5100\AA}$=1.60$\times$10$^{44}$ erg s$^{-1}$ corresponding to 
a corrected SMBH mass of $\sim$3.2$\times$10$^{7}$ M$_{\sun}$.

The value of M$_{BH}$ reported by \citet{Shen2011} for \src\ is 3.8$\times$10$^{7}$. This
value is not corrected for the host galaxy contamination (nor for the extinction, apart from the Galactic
one) but an empirical formula  is provided (their eq.~1) to estimate the strenght of this 
contamination at 5100\AA. 
If we use this formula to correct the M$_{BH}$ reported by \citet{Shen2011} 
we obtain M$_{BH}$=3.2$\times$10$^{7}$ M$_{\sun}$ which is in  
very good agreement with our estimate. 
The value of SMBH mass derived for \src\ 
is in the typical range of masses observed in the RL NLS1 
(\citealt{Komossa2006, Doi2012, Foschini2012a}).

Using the bolometric correction
(BC$_{5100}$=10.33) given by \citet{Richards2006} to compute the bolometric 
luminosity starting from the luminosity
at 5100\AA\ (corrected for the host galaxy contamination) we obtain 
L$_{bol}$=1.65$\times$10$^{45}$ erg s$^{-1}$ and an Eddington ratio of 
0.40.
In the estimate of the bolometric correction \citet{Richards2006}
consider the integrated emission of the AGN composite spectrum from
100$\,\mu$m to 10keV thus including the infrared part of the SED. 
In the standard view of an AGN the IR emission is usually attributed to
the reprocessed emission of the accretion disk and, therefore, it is often
excluded from the estimate of the real energetic output of the source. 
Since the IR hump usually represents about 1/3 of the total emission 
(\citealt{Marconi2004}), a more reliable estimate of the actual bolometric luminosity of 
\src\ is $\sim$1.1$\times$10$^{45}$ erg s$^{-1}$ corresponding to 
an Eddington ratio of $\sim$0.27.
The high accretion rate, close to the Eddington 
limit, observed in  \src\ is, again, typical of the NLS1 class. 

As previously mentioned, the SMBH mass derived from the virial method,
like the single epoch relation mentioned above, can be significantly 
underestimated in NLS1 if the BLR is disk-like (e.g. \citealt{Decarli2008}) 
or if the radiation pressure (important
at high Eddington ratios) affects the kinematics of the BLR 
(\citealt{Marconi2008}). We use the relation 
presented in \citet{Chiaberge2011} to estimate the SMBH
mass once the radiation pressure is taken into account. We find a
factor $\sim$2 (0.3 dex) larger value of mass 
(M$_{BH}^{Prad}$=6.6$\times$10$^7$ M$_{\sun}$)
and a consequently lower Eddington ratio (0.13).
The actual importance of the radiation pressure, however, is still
matter of debate (e.g. see \citealt{Netzer2009}).

Independent estimates of the SMBH mass can be derived using the link between 
the SMBH mass and the stellar velocity dispersion 
(\citealt{Gultekin2009}):

\begin{equation}
Log\left(\frac{M_{BH}}{M_{\sun}}\right)= 8.12+ Log \left(\frac{\sigma}{200 km s^{-1}}\right)^{4.24}
\end{equation}

A rough estimate of $\sigma$ can be inferred from the width of the narrow 
lines, like the [OIII]$\lambda$5007\AA, assuming that the NLR is influenced 
by the potential of the galaxy and that the gas follows the same kinematics 
as stars. Using a sample of type~2 AGN selected from the SDSS,  
\citet{Greene2005} have shown that, indeed, narrow lines like the 
[OIII]$\lambda$5007\AA\ can be used as a proxy for the dispersion velocity
of the bulge provided that the blue wings are correctly modelled and removed.
\citet{Komossa2007} and \citet{Komossa2008}  
have noted that the [OIII]$\lambda$5007\AA\ 
width can be used as a proxy for the stellar velocity only in those 
NLS1 that do not show large blue-shifts\footnote{These 
offsets are computed 
relative to the 
[SII]$\lambda\lambda$6716,6731\AA\ emission lines} (hundreds of km s$^{-1}$, 
e.g. \citealt{Marziani2003}) in the ``core'' of the [OIII] line 
(i.e. excluding the wings). 
These blue-shifts suggest that this line is unsuitable 
for tracing the stellar velocity in the bulge. 
In the case of \src\ we observe blue wings, as described in the
previous section, but the core of the line does not show any significant 
offset ($<$50 km s$^{-1}$) 
with respect to the [SII]$\lambda\lambda$6716,6731\AA\ emission lines. 
Therefore, the  [OIII]$\lambda$5007\AA\ width can be 
considered
as a reasonably good proxy of the stellar velocity. 
The [OIII]$\lambda$5007\AA\ width (excluding the blue wings
), 
corrected for the instrumental resolution, is 
$\sigma$=FWHM/2.355=145 km s$^{-1}$ which leads to a SMBH mass of 
3.4$\times$10$^{7}$ M$_{\sun}$, with a statistical error of $\sim$6 per cent
but with a much larger uncertainity due to the intrinsic scatter on the 
M-$\sigma$ relation (0.44 dex, \citealt{Gultekin2009}). This  
value of mass is very close to the one obtained directly 
from the virial method without taking into account
the radiation pressure.
Given the large errors  involved in all the methods used to derive the SMBH 
mass it is not possible to test and quantify the importance of the 
radiation pressure using only one source but we consider the results presented above 
as an indication that the possible systematics related to the orientation of
the source or to the radiation pressure are well within the expected uncertainty
related to the SE virial method.

In conclusion, our best estimate of the mass of the central SMBH of 
\src\ is 3.2$\times$10$^{7}$ M$_{\sun}$ with an uncertainty (associated to the
SE virial method) of $\sim$0.4 dex.

\section{Radio emission}

\subsection{Radio loudness parameter}
SDSSJ143244.91+301435.3 is detected in many of the existing radio surveys (see 
Tab.~\ref{tab_radio}): 
the NVSS (\citealt{Condon1998}) and FIRST (\citealt{Becker1995}), at 1.4~GHz, 
the WENSS survey (\citealt{Rengelink1997}) at 
325~MHz and the 7C (\citealt{Hales2007}) at 151~MHz. The source is not 
present in the VLA Low-Frequency Sky Survey Redux (VLSSr) catalogue at 74~MHz (\citealt{Lane2012})
and we have estimated an upper limit of $\sim$600~mJy.
The source is not detected 
at higher frequencies (5~GHz) in the GB6 catalogue (\citealt{Gregory1996}).
In order to have information also at these frequencies we have observed 
\src\ with the Effelsberg 100-m radio telescope
at 2.64, 4.85, 8.35 and 10.45~GHz (see the observing log in
Tab.~\ref{eff_data_log}). The source has been 
detected at all frequencies.

A rough indicator of the level of radio-loudness of the source is the 
radio-loudness parameter defined as in \citet{Kellermann1989}:

\begin{equation}
R = \frac{S_{5~GHz}}{S_{4400\AA}}
\end{equation}

where the S$_{5~GHz}$ is the observed flux density at 5~GHz 
and S$_{4400\AA}$ is derived from the B magnitude 
(S$_{4400\AA}$=4.53$\times$10$^{-0.4B+6}$ mJy, according to 
\citealt{Kellermann1989}). 
Using the SDSS g and u magnitudes to estimate the
B magnitude (\citealt{Jester2005}): 

\begin{equation}
B=g+0.17\times(u-g)+0.11
\end{equation}

we obtain R$\sim$160 a value well above the radio-loud/radio-quiet 
dividing line (R=10) proposed by \citet{Kellermann1989}. 

This value of R is useful for a comparison with previous works but it is just a
crude approximation of the actual radio-to-optical flux density ratio of the
source since it is based on observed flux densities, i.e. not k-corrected.
To have a more accurate number and for an easier comparison with recent
samples of RL NLS1, we have estimated the value of 
R$_{1.4}$ at 1.4~GHz according to \citet{Yuan2008}:

\begin{equation}
R_{1.4} = \frac{S_{1.4~GHz}}{S_{4400\AA}}
\end{equation}

where the flux densities are referred to the source rest-frame.
Using the k-corrected\footnote{For the radio we use a spectral index of 
$\alpha$=0.93 (see next section) while for the optical flux density we use the
optical spectrum to derive the 4400\AA\ flux density (rest frame)}  radio and 
optical flux densities we obtain R$_{1.4}$=560. 
Being R$_{1.4}$ defined at lower radio frequency than R its
value is usually larger than R. 

\begin{center}
\begin{table}
\caption{Radio data-log of the Effelsberg observations}
\label{eff_data_log}
\begin{tabular}{l r r r r}
\hline\hline
 Date       & Freq.    &  Bandwidth & Beam size	 & Flux density\\
            &          &            &            &             \\
            &  (GHz)   &   (MHz)    & (arcsec)   &   (mJy)     \\
\hline
22.07.2013 & 4.85  & 500    & 145  & 18.3$\pm$2.0 \\
         & 8.35    & 1100   &  89  & 8.0$\pm$1.0 \\
          &10.45   & 300    &  68  & 5.4$\pm$1.0 \\
30.09.2013   & 2.64 & 80    & 276  & 38.0$\pm$2.0 \\
           &10.45   & 300    &  68 & 7.3$\pm$1.2 \\ 

\hline
\end{tabular}
\end{table}
\end{center}
}

We note that the flux density at 4400\AA\ could be in part contaminated by the
host galaxy and, at the same time, it can be affected by the 
extinction. The two effects act in the opposite direction, i.e. 
the presence of the host galaxy increases the observed flux density while
the extinction reduces its value. 
Similarly to what we have done in the estimate of the mass of the
SMBH, we can use the results of the SED modelling described in Section~3 
to evaluate the two effects. 
Applying all the corrections and considering the uncertainties on
these estimates we obtain R$_{1.4~GHz}\sim$600$\pm$100.

   \begin{figure*}
   \centering
    \includegraphics[width=16cm]{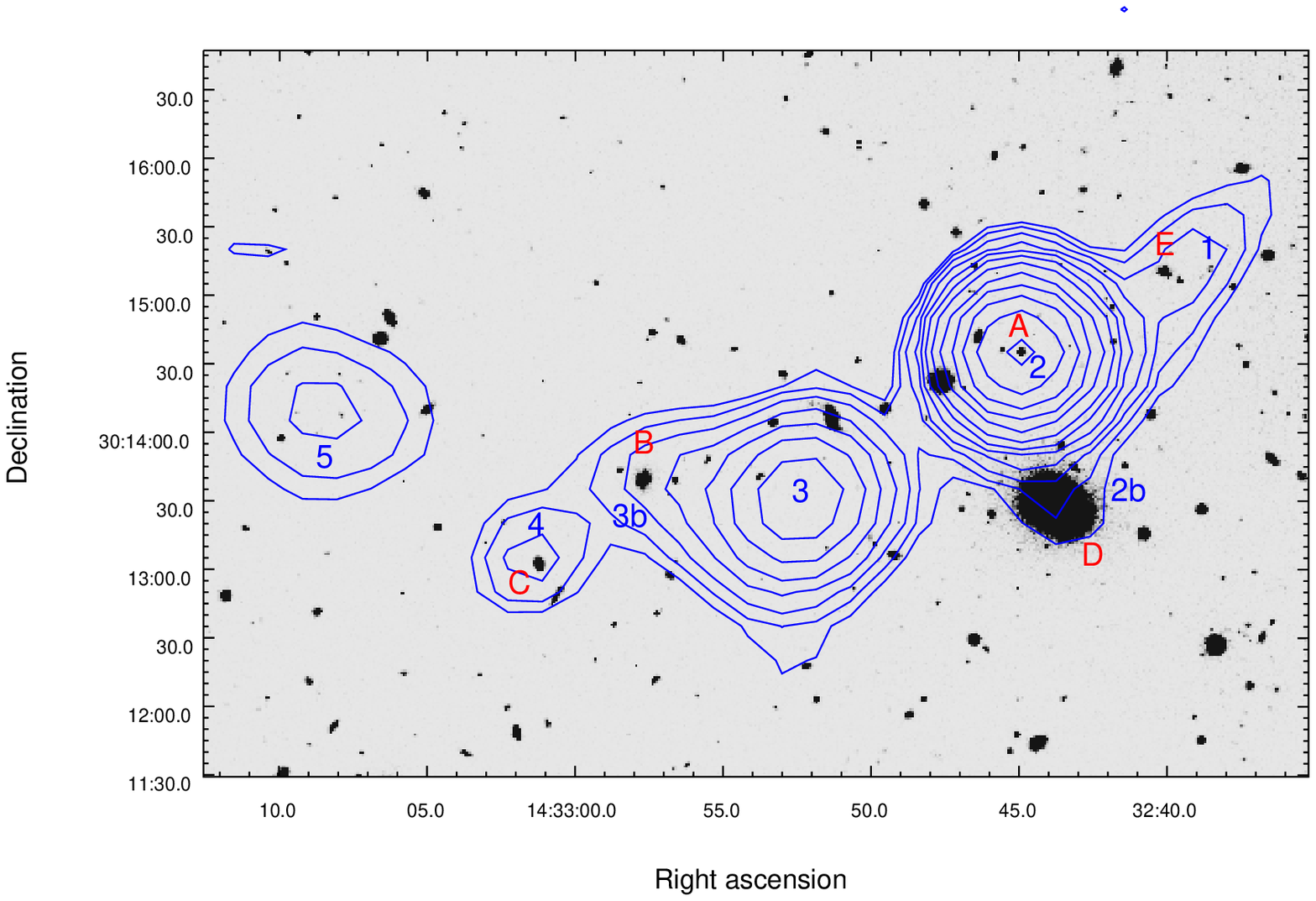}
    \includegraphics[width=16cm]{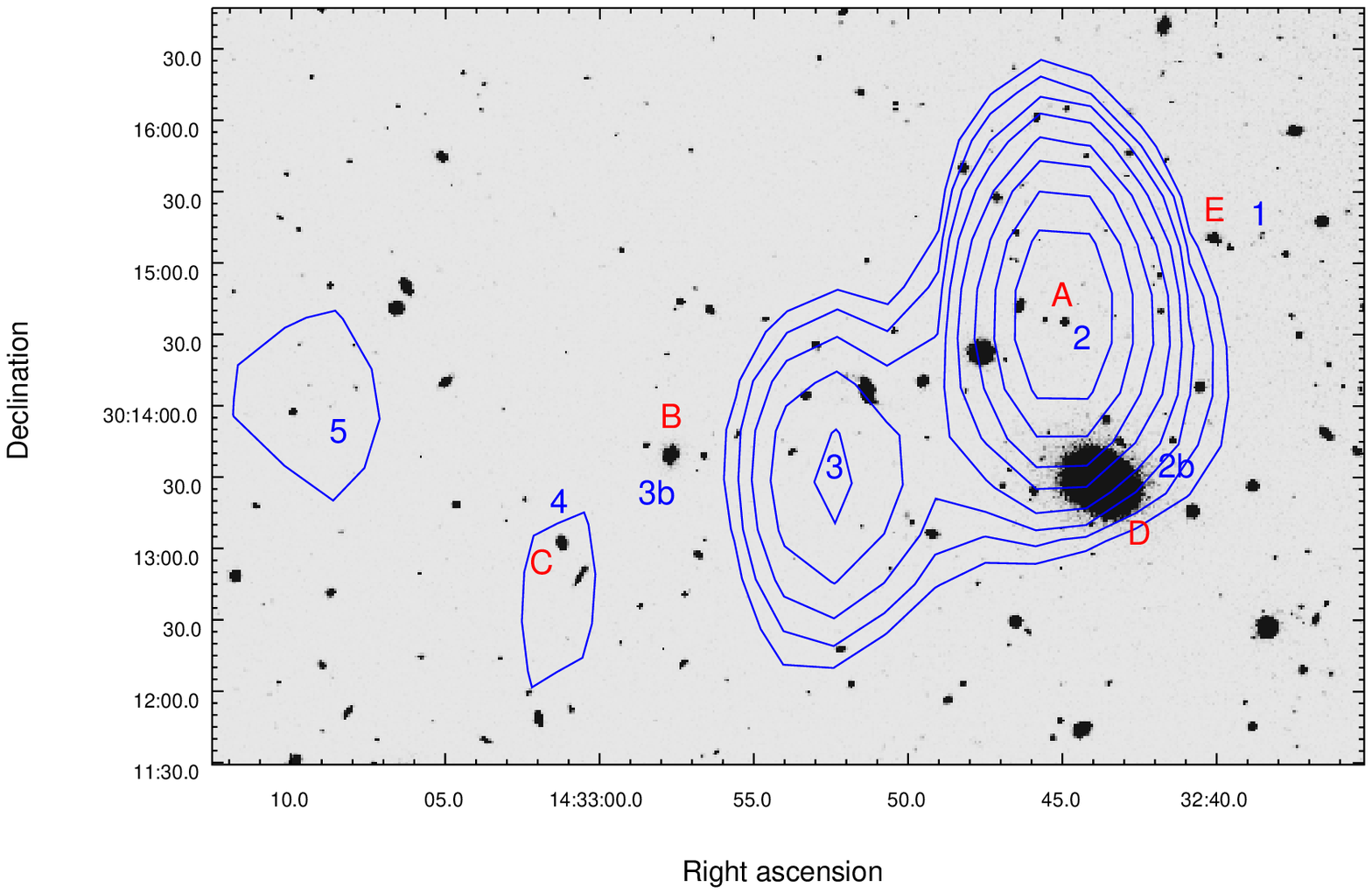}
   \caption{SDSS image in the r-filter with, superimposed, the radio contour levels from NVSS 
(1.4~GHz, upper panel) and WENSS (325~MHz, lower panel). Contour levels start from 1~mJy (NVSS) and 10~mJy (WENSS)
and increase by multiplicative steps of $\sqrt 2$. 
The numbers indicate the radio components detected at 1.4~GHz by NVSS (labels 2b and 3b indicate additional 
components found by FIRST) while letters  indicate the optical counterpart candidates. The NLS1 is 
object A (which is positionally coincident with the radio source 2).
North is up, east to the left}
              \label{image_r_radio}
    \end{figure*}

The value of R$_{1.4}$ clearly puts this source among the most radio-loud
NLS1 discovered so far. To date, only 13 RL NLS1 with 
R$_{1.4}\geq$500 have been discovered (\citealt{Yuan2008, Foschini2011}).

\begin{table*}
\caption{Radio sources in the field of \src. 
Numbers indicate the radio components detected in the NVSS catalogue while a letter after the number 
indicates a sub-component detected at higher resolution in the FIRST catalogue.}
\label{tab_radio}
\begin{tabular}{c c c c c c c c c c c c}
\hline\hline
Comp. & S$_{10.45~GHz}$ & S$_{8.35 GHz}$ & S$_{4.85 GHz}$ & S$_{2.64 GHz}$ & S$_{1.4 GHz}^{peak}$ & S$_{1.4 GHz}^{int}$ & S$_{1.4 GHz}^{peak}$ & S$_{1.4 GHz}^{int}$ & S$_{325 MHz}^{peak}$ & S$_{325 MHz}^{int}$ & S$_{151 MHz}$ \\ 
                &    (Eff.) &  (Eff.) &  (Eff.) & (Eff.) & (NVSS)               &     (NVSS)            &      (FIRST)          &   (FIRST)             &    (WENSS)              &   (WENSS)  & (7C)    \\
                &    mJy    &   mJy   & mJy   & mJy & mJy/beam            &      mJy              &       mJy/beam        &    mJy                &     mJy/beam            &     mJy    &  mJy    \\
\hline
1            & - & - & - & -  &  2.3                   &   5.9                 & 1.27                  & 1.74                  &  -                   &  -    &  -    \\
2             & 6.4$^1$ & 8.0 & 18.3 (11.7)$^2$ & 38.0 (27.6)$^2$  &  50.6        & 50.6        & 49.94                 & 49.98                 &  166                    & 195      &  330    \\
2b            & - & - & - & -   &    -                  &    -                  & 0.68                  &  1.04                 &   -                    &  -      &  -       \\
3            & - & - & -  & -   &   10.6                 & 18.9                  & -                  &  -                 &   45                    &  51      &  -    \\
3b          & - & - & -  & -    &    -                  &    -                  & 2.0                   & 1.4                   &   -                    &  -      &  -     \\ 
4           & - & - & -  & -   &   2.7                  &  2.7                  & 2.49                  & 2.79                  & -                    & -     &  -       \\
5         & - & - & -  & -   &   3.5                  &  3.5                  & -                  & -                  & -                    & -     &  -       \\

\hline
\end{tabular}
\hfill \break
$^1$ this is the average value of the two flux densities measured in July and September (see data-log in Table~\ref{eff_data_log}).
\hfill \break
$^2$ the numbers between parenthesis are the flux densities corrected for confusion (contamination from source~3)
\hfill \break
\end{table*}

\subsection{Radio morphology and possible components}

The integrated flux density at 1.4~GHz is 50.6 mJy, in the NVSS, and 50.0 mJy, 
in the FIRST.
The source is barely resolved in the FIRST survey which has a higher 
angular resolution than NVSS. In the FIRST catalogue the source has a 
de-convolved size of 0.28$\arcsec$ (major axis) that should be considered
as an upper limit of the source size. This limit corresponds to a linear 
size of 1.4~kpc. 

At low-frequency (325~MHz), the source is detected with a peak flux density of 
166~mJy/beam and an integrated flux density of 195~mJy. The discrepancy 
between the two flux densities (of $\sim$17 per cent) 
may indicate a partial extension at this frequency.

In addition to the radio emission centred on SDSSJ143244.91+301435.3, 
the NVSS map shows many other radio 
sources/components in the relatively nearby sky region ($\sim$10$\arcmin$) as shown in 
Fig.~\ref{image_r_radio}.
Apart from 
the radio source associated to SDSSJ143244.91+301435.3 (component 2) some of these 
additional radio sources are clearly associated to optical objects 
(components 1 and 4) while other 
emissions (3 and 5) do not seem to be obviously related to an optical counterpart. In particular, the emission
labeled as ``3'' is quite strong ($\sim$18.9 mJy of integrated flux density) and relatively close 
($\sim$2$\arcmin$) to SDSSJ143244.91+301435.3. Contrary to the radio source coincident 
with SDSSJ143244.91+301435.3,
source 3 is clearly extended even at the NVSS resolution (beam size of $\sim$45$\arcsec$) 
and it is not detected in the FIRST survey. A possible 
hypothesis is that source 3 is a radio-lobe associated to SDSSJ143244.91+301435.3 (object 
A\footnote{We use numbers to indicate the radio components and letters to indicate 
the optical sources}, in the figure). 
The linear distance
between the core and the radio lobe (measured from the peak emission) would be $\sim$580 kpc and the source  
would appear very asymmetric with just one-sided radio lobe. 

A second (more likely) possibility, however, is that source 3 is associated to the optical object labeled 
as B (an early type galaxy at z=0.2075) even if not positionally coincident. 
This is suggested by the fact that source 3 
is elongated towards object B.
In the FIRST survey this ``elongation'' is detected as a faint source (component 3b in Tab.~\ref{tab_radio},
S$_{1.4 GHz}\sim$2 mJy)
positionally coincident with object B (see Fig.~\ref{image_r_radio}). It is therefore possible that 
source 3 is a radio lobe produced by galaxy B rather than SDSSJ143244.91+301435.3. In this case, source 5, 
another radio source not obviously associated to an optical counterpart, could be the second radio lobe on the
opposite side. Under this hypothesis, the linear distance between core and lobe (source 3) 
is $\sim$230 kpc while the total size of the system (i.e. the distance 
between the two lobes) is $\sim$720 kpc. All these parameters are perfectly reasonable for a
radiogalaxy. Also the optical spectral type of object B, an early type galaxy, is consistent
with what is observed in many radiogalaxies.  

At this stage it is difficult to establish on a firm ground which of the two hypotheses described above is
correct. More radio data are required to detect, for instance, any possible
``bridge'' between the radio lobe candidate and one of the two possible radio cores. However, we consider 
as more probable the hypothesis that the radio component 3 is not associated to \src\ since, otherwise, 
the system would be extremely asymmetric with just one bright lobe on one side. For this reason in the
following discussion we will assume that only the radio emission labeled as 2 is associated
to \src\ (this association is certain).

\begin{center}
\begin{table*}
\caption{Properties of the possible optical counterparts}
\label{tab_opt}
\begin{tabular}{c l c c c c c c c}
\hline\hline
name            & position                    & u     	  & g	          &  r	          & i   	  & z            & redshift     & comments \\ 
                &  (J2000)                    & (SDSS)    &   (SDSS)      & (SDSS)        &   (SDSS)      &  (SDSS)      &          \\
                &                             &           &               &               &               &              &          \\
\hline
A               & 14 32 44.91 +30 14 35.4    & 19.14     &	18.82     &	18.56     &	18.62     &	18.11    &0.355        & NLS1  \\
B               & 14 32 57.68 +30 13 39.5    &	20.31     &	18.83     &	17.53     &	16.96     &	16.59    &  0.2075      & Early type galaxy  \\
C               & 14 33 01.22 +30 13 02.4    & 19.45	  &     18.53	  &     18.04	  &     17.66     & 	17.55    & -           & -   \\
D               & 14 32 43.90 +30 13 29.2    &	17.34     & 	15.40     & 	14.52     &	14.10     &	13.79    & 0.0623      & Early type galaxy \\
E               & 14 32 40.14 +30 15 10.5    & 21.19     &	19.69     &	19.14     &	18.85	  &     18.75    & -           & - \\
\hline
\end{tabular}
\end{table*}

\end {center}

\subsection{Radio spectrum}

\begin{figure}
   \centering
    \includegraphics[width=8.5cm]{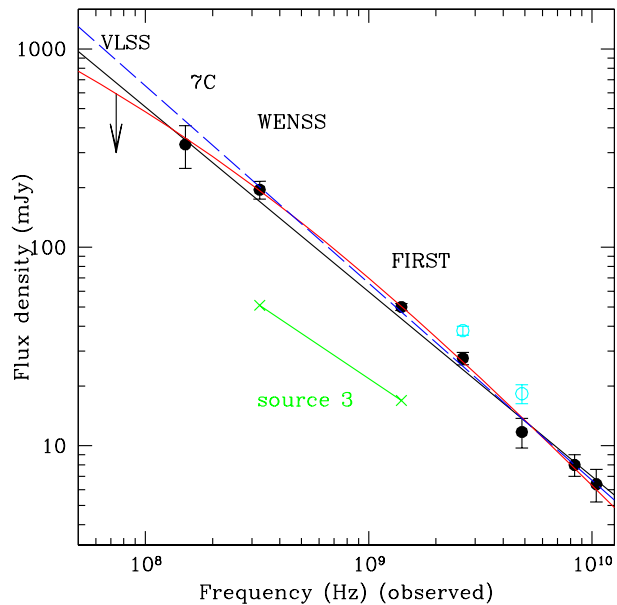}

   \caption{Radio spectrum of \src. High-frequency data ($>$2.6~GHz) come from 
Effelsberg observations while lower-frequency data come from existing 
catalogues (see labels). 
The 2 open points (cyan in the electronic version) are the Effelsberg 
flux densities at 2.64 and 4.85~GHz (where the beam size is larger than 
2$\arcmin$) not corrected for the contamination of the nearby source~3 (whose 
spectrum is also plotted with the two green crosses and line).
The two fits to the points using a single power-law and a smooth broken 
power-law (with a break at 300~MHz) respectively are also plotted as solid 
lines (black and red, respectively, in the 
electronic version). Finally, the dashed (blue) line is the fit to the 
points at the highest frequencies ($>$1 GHz).}
              \label{radio_spect}
    \end{figure}

All radio flux densities available/measured for \src\ are reported in Tab.~\ref{tab_radio} (source 2) and
plotted in Fig~\ref{radio_spect}. A least square  fit to the data 
results in a quite steep ($\alpha\sim$0.9) 
spectrum. Since, as discussed above, this region of sky is quite crowded (see Fig.~\ref{image_r_radio}) 
it is possible that some of the measured flux densities, in particular those taken with 
a ``single dish'' (Effelsberg), are confused by nearby sources. 
Among all sources which are close enough to \src\ to be potentially contaminating the observed
fluxes, two are quite faint (sources 1 and 2b in Tab.~\ref{tab_radio}).  
Their possible contamination of the flux density of \src\ is expected to be below $\sim$10 per cent.
The principal contamination is expected to come from  source 3
which is relatively strong (18.9 mJy). Given its distance from
\src\ ($\sim$2$\arcmin$) any flux density measured with a beam size $\geq$2$\arcmin$ is contaminated.
The confusion problem, therefore, is likely present in the Effelsberg measurements at the 
two lowest frequencies (4.85~GHz and 2.64~GHz) where the beam size is 2.4$\arcmin$ and 
4.6$\arcmin$ respectively, while the flux densities at 8.35 and 10.45 GHz should be
relatively free from contamination (beam size of 1.5$\arcmin$ and 1.1$\arcmin$ respectively).
We have estimated the contribution of source~3 to the flux density 
of \src\ (at 4.85~GHz and 2.64~GHz) using the
observed NVSS integrated flux density of source~3 at 1.4~GHz ($\sim$17~mJy, excluding the contribution from 
source~3b) and 
extrapolating it to the 2 frequencies using the observed spectral index of this source 
(see Fig~\ref{radio_spect}). 
The corrected flux densities are more aligned
with the points at lower and higher frequencies, confirming that the contamination from source 3
is actually present at 4.85~GHz and 2.64~GHz at the level estimated from the NVSS data. 
The fit
to the data corrected for the contamination does not change significantly leading to a value of the
slope close to the previously measured one ($\alpha$=0.93).

Most of the radio data used in the analysis were taken at 
different epochs. In principle, variability may play a role 
and influence the computed spectral index. 
However, we have several indications that variability in \src\ 
is not important: FIRST and NVSS data (taken in April
1993 and April 1995, respectively) are fully consistent (see 
Tab.~\ref{tab_radio}) 
and the 10.45~GHz flux densities taken at Effelsberg in 
two different sessions (July and September 2012) are consistent with a 
constant flux density within $\sim$ 1 $\sigma$ (see Tab.~\ref{eff_data_log}). 
Moreover, we have fitted Effelsberg data taken at the same epoch obtaining 
very similar slopes (0.93 and 0.97).
We conclude that current data do not suggest the presence of radio variability 
in \src.

Although a simple power-law offers a reasonable description of the spectrum, the measurements at the
lowest frequencies seem to suggest a possible flattening of the slope. If we fit only
the flux densities at the highest frequencies ($>$1 GHz) and extrapolate the fit to lower frequencies, 
hints for a possible break at $\sim$300 MHz seem to emerge (Fig.~\ref{radio_spect}). Using a broken power-law model 
we obtain a good fit with a break frequency at $\sim$300~MHz (corresponding
to 400~MHz, rest frame). 
The spectral indices above and below the
break are 1.2 and 0.5, respectively. On the basis of the available 
data, however, the  presence of a turnover at low-frequencies is not statistically 
significant and it should be considered just as a
possible indication. 


\subsection{\src\ as CSS source}
Most of the RL NLS1 discovered so far with radio-loudness parameters similar
to the one computed in \src\ present flatter ($\alpha<0.5$) radio spectra and
show indications for the presence of relativistic beaming, something that
suggests the presence of a relativistic jet pointing towards the observer 
(\citealt{Yuan2008}). For comparison, in the \citet{Yuan2008} sample of very radio-loud
NLS1 only
one source (J1443+4725), among those with a measured spectral index, has
a relatively steep ($\alpha$=0.67) radio spectrum. In the \citet{Komossa2006}
sample there are  2 sources (SDSS J172206+565451 and RX J0134--4258)
with a high value of radio-loudness and having a steep (0.69-1.4) 
radio spectrum\footnote{The source TEX~11111+329 in the \citet{Komossa2006} 
sample
has a very steep spectrum (1.24 between 1.4 and 5~GHz) but the radio-loudness
parameter, once corrected for the high level of reddening measured in this
object, is $\sim$20 i.e. much lower than the one observed in \src}.
The steep radio spectrum observed in \src\ 
seem to exclude that we are observing a jet 
pointing in our direction. This is further supported by the non-detection
of a strong flux variability, as explained earlier.
Therefore, \src\ is one of the very few RL NLS1 with a high radio-loudness 
parameter that could belong to the parent population.

With the observed radio properties the source can be classified as
a CSS radio sources (\citealt{O'Dea1998, 
Dallacasa1995}). 
The radio power of \src\ (2.1$\times$10$^{25}$ W Hz$^{-1}$ 
at 1.4 GHz) is in the range usually 
observed in CSS sources (e.g. \citealt{O'Dea1998, Fanti2001}). CSS 
sources have a linear size between 1 and 20 kpc and present a spectral 
turnover at
low ($<$500 MHz) frequencies attributed to synchrotron self-absorption or 
free-free absorption. The position of the turnover is inversely
correlated with the linear size of the source and, for sources with 
a linear size of $\sim$1.4 kpc, we expect a turnover at $\sim$100-500~MHz 
(e.g. \citealt{O'Dea1998}).
As previously discussed there is a possible hint (although not statistically 
significant) of a change of slope at low-frequency (observed at $\sim$300~MHz
corresponding to 400~MHz, rest frame) that may 
indicate a possible 
turnover of the spectrum at a frequency close to the
one expected.
The observed radio spectral index (0.93 considering a single power-law 
fit, or 1.2,
considering the broken power-law fit) is within the distribution
of indices, measured above the spectral turnover, of the CSS sources (between 
0.6 
and 1.2, \citealt{Fanti1990, O'Dea1998}). Finally, in the NVSS data
no significant polarization is found ($\leq$1 per cent) something that, again,
is consistent with what is observed, at these frequencies, in CSS sources with 
linear size below $\sim$6 kpc (\citealt{Cotton2003}).

CSS sources are compact systems of sub-galactic sizes that 
show radio morphologies  very similar 
to those observed in ``classical'' radio galaxies but on much smaller scales. 
Interestingly, many CSS sources show
distorted morphologies suggestive of interactions with a dense and 
inhomogeneous 
environment (\citealt{Dallacasa2013, O'Dea1998}).
At present, we have no high resolution radio images of \src\ that
could confirm, also from the morphological point of view, that \src\ 
is a true CSS source but, given its steep radio spectrum, 
it is very likely that the source would appear resolved at VLBI scales.

\subsection{Radio emission and SF}
In the SED modelling discussed in Section~3 the IR photometric
points play an important role to detect and characterize the host galaxy
properties. The 12 and 22$\,\mu$m points (corresponding to 8.9 and 
16.2$\,\mu$m, rest frame), in particular, are critical to 
establish the possible starburst nature of the host. We have then
shown that \src\ presents a strong radio emission that is likely
ascribed to the presence of a jet.
In principle, the possible presence of a jet could contaminate the observed
IR emission leading to incorrect conclusions about the presence of SF in 
the host galaxy.
For this reason, we want to assess the level of this possible contamination in the 
1-20$\,\mu$m spectral region. To this end, we have to assume a radio-to-IR
spectral slope. One possibility is  to simply extrapolate the slope 
($\alpha\sim$0.9)
observed in the radio band up to the IR region. With this slope we predict 
a flux density at 22$\,\mu$m which is 3 orders of magnitude lower than the one 
observed. 
Nevertheless, the radio spectrum is based on the observed emission
which is likely dominated by extended components, like mini-lobes, that
have steep spectra. The nuclear emission, instead, is expected to have a flatter 
spectrum, more similar to the one observed in blazars. For this reason, we
have also considered the average radio-to-IR slopes derived from the 
Caltech-Jodrell Bank Flat spectrum (CJ-F, \citealt{Taylor1996}) sample of 
blazars. 
This is a flux limited and
complete sample of flat spectrum radio sources with a radio flux density
greater than 350 mJy at 5~GHz. The SED of most of the 293
CJ-F blazars is under study (Ant\'{o}n et al. in prep.).
Using the average value of slope between 5~GHz and 22$\,\mu$m of  
the CJ-F blazars ($\alpha_{5GHz}^{22\,\mu m}$=0.64) we
predict a flux density which is 2 orders of magnitude below the observed value. 
Even considering the rms uncertainty on the value of this slope
we still predict a contribution of jet to the 22$\,\mu$m flux density which is
less than 10 per cent.
We conclude that the possible contamination of the jet emission to the
observed IR emission is not expected to be significant. 
The presence of a significant SF is therefore the
best explanation for the observed IR emission at $\sim$10-20$\,\mu$m.  

We finally note that a high SF is  expected to produce an important
radio emission that can represent a relevant contribution to the observed 
radio luminosity of \src. 
If this contribution was significant it could explain the steep radio spectrum observed
in \src.
We can estimate the expected radio emission due to the SFR 
from the following equation (\citealt{Condon2002}):

\begin{equation}
\frac{P_{1.4 GHz}}{W Hz^{-1}} \sim 4.6\times10^{21} \frac{SFR (M>5 M_{\sun})}{M_{\sun} y^{-1}}
\end{equation}

\noindent
Using a SFR of 50 M$_{\sun}$ y$^{-1}$ we predict a radio luminosity of 2.3$\times$10$^{23}$ W Hz$^{-1}$
which is a factor $\sim$100 lower than the observed P$_{1.4 GHz}$. 
A similar value (1.5$\times$10$^{23}$ W Hz$^{-1}$) is obtained if we use the radio vs FIR luminosity
relation computed for the IRAS Bright Galaxy Sample presented in \citet{Condon2002}. 
We conclude that the observed radio emission should be only marginally ($\sim$1 per cent) contaminated by the SF in the host galaxy.

\section{High-energy emission}
\subsection{X-rays}
X-rays can provide potentially important pieces of information on \src. At these energies we 
expect to observe
the nuclear emission typical of RQ AGN, which is likely produced by a hot
corona of electrons reprocessing the primary UV-optical emission of the 
disk via inverse-Compton mechanism (\citealt{Haardt1991}). In RQ NLS1, in particular, the X-ray 
emission is often characterized by steeper 2-10 keV slopes, when compared to BLS1 
(e.g. \citealt{Brandt1997, Zhou2010, Grupe2010, Caccianiga2011b}), and by 
the presence of a strong and variable soft-excess at lower energies ($<$2 keV, \citealt{Boller1996}). 
XMM-{\it Newton} observations have also revealed an interesting 
spectral complexity (spectral drops/curvature) in many NLS1 (e.g
\citealt{Gallo2006a} and references therein) that has been 
interpreted as caused either by the presence of dense material in the 
close environment of the central SMBH (``partial covering model'', 
e.g. \citealt{Tanaka2004}) or by the reflection of the primary emission
on the accretion disk (``reflection model'', e.g. \citealt{Fabian2004}).

Besides the coronal emission, we also expect a contribution from the relativistic jet, similar to 
that observed in blazars. The relative intensity of the RQ and RL components strongly depends 
on the level of beaming present in the source. In ``beamed'' RL NLS1 the emission from the jet could be
very important or even dominant  (e.g. \citealt{Gallo2006}) while
in a ``mis-oriented'' source like \src\ the jet emission is probably less relevant.  
In addition, since \src\ is also a possible CSS source, we can expect  X-ray emission from
the shocked heated IGM caused by the expanding radio jet (e.g. \citealt{Siemiginowska2009}). 
Finally, the presence of a quite intense SF can also contribute to the total budget of the 
X-ray emission. 

We have searched for X-ray data from the existing surveys but no detection was found. We can derive
an upper limit on the soft (0.1-2.4 keV) X-ray emission of \src\ from the ROSAT all sky survey (\citealt{Voges1999}) of
$\sim$5$\times$10$^{-13}$ erg s$^{-1}$ cm$^{-2}$ and an upper limit in the 2-12 keV band from the
XMM-{\it Newton} Slew Survey (XSS, \citealt{Warwick2012}) of $\sim$6$\times$10$^{-12}$ erg s$^{-1}$ cm$^{-2}$ 
(assuming a photon index of 2). These upper limits correspond to limits on the X-ray 
luminosity of $\sim$2$\times$10$^{44}$ erg s$^{-1}$ and $\sim$3$\times$10$^{45}$ erg s$^{-1}$ in the
0.1-2.4 keV and 2-12 keV energy bands, respectively.
In order to evaluate whether these limits are low enough to set some useful constraint on the
emission of \src, we estimated the predicted  X-ray luminosities from the two main components
expected from this object, i.e. the coronal and the jet emission. For what concerns the first
component, we start from the bolometric luminosity, derived in the previous sections, combined to 
the value of X-ray bolometric correction\footnote{K$\rm_{bol}$ is defined as the fraction between the bolometric luminosity 
and the X-ray luminosity in the 2-10 keV band} typically
observed in AGN accreting with an Eddington ratio of 0.27, using the
relations between K$\rm_{bol}$ and Eddington ratio recently published in \citet{Fanali2013a}. 
For the jet component, we use the
observed radio power at 5~GHz and assumed the mean value of the radio-to-X-ray luminosity ratio 
typically observed in flat-spectrum radio quasars (FSRQ, \citealt{Donato2001}).
The expected values of X-ray luminosity in the 0.1-2.4 keV (2-10 keV) energy band are 
$\sim$7$\times$10$^{43}$ erg s$^{-1}$ 
($\sim$4$\times$10$^{43}$ erg s$^{-1}$) and $\sim$1$\times$10$^{43}$ erg s$^{-1}$ 
($\sim$6$\times$10$^{42}$ erg s$^{-1}$) 
for the coronal and the jet emission, respectively.
The emission predicted from the jet should be considered only as an upper limit since most of
the radio power observed at 5~GHz is expected to come from extended structures, as previously
discussed, and not from the radio core. Both estimates of the X-ray luminosity are more than a factor
$\sim$100 below the upper limits derived from the XSS and more than a factor $\sim$3 
below the RASS limit and, therefore, the null detection of \src\ is not surprising.  
In addition, it should be considered that 
NLS1 are variable X-ray sources that are sometimes observed in very low flux states 
(e.g. \citealt{Grupe2012}) and this may have contributed to the null
detection of \src\ in the existing X-ray surveys.
Pointed observations with one of the existing X-ray telescopes should be able to detect \src\ 
and to provide enough counts to carry out a reliable spectral analysis.

\subsection{Gamma-rays}
An interesting issue related to RL NLS1 is that they have been often found 
to be gamma-ray emitters, like the class of blazars and radiogalaxies. 
After the first detection
of a NLS1 (PMN~J0948$+$0022) in the {\it Fermi}-LAT catalogue (\citealt{Abdo2009, Foschini2010}) 
more cases have been discovered (\citealt{Abdo2009a}) and the number of RL NLS1 detected in 
the {\it Fermi}-LAT
catalogue keeps on increasing (see \citealt{Foschini2011, Foschini2013d}). Contrary to
many of the RL NLS1 discovered so far, the radio properties of \src, 
in particular its steep radio spectrum and the lack of polarization and significant  
variability, 
suggest that most of the observed radio emission should come from extended
structures, like mini-lobes, and not from the core. Therefore it is reasonable
to expect that any gamma-ray emission in this object 
should be lower than in blazars or in core-dominated RL NLS1 with a similar 
observed radio luminosity. Indeed, so far, {\it Fermi}-LAT has preferentially detected
blazars or radio-galaxies characterized by large core-dominance values
(\citealt{Abdo2010}) with the only exception of two very nearby radio galaxies
(Cen~A and M87). Also considering \src\ as a ``young'' CSS source, its detection by 
{\it Fermi}-LAT detection can have important implications, as recently 
discussed by \citet{Migliori2013}.

We have thus analysed the {\it Fermi}-LAT public 
data\footnote{http://fermi.gsfc.nasa.gov/cgi-bin/ssc/LAT/LATDataQuery.cgi} searching 
for  a detection, or an upper limit, of
the gamma-ray emission of \src.
{\it Fermi}-LAT data between August 4, 2008 and August 4, 2013 (5 years) were
retrieved from HEASARC and analyzed using the standard
procedures\footnote{http://fermi.gsfc.nasa.gov/ssc/data/analysis/scitools/}.
Specifically, we used LAT Science Tools v. 9.27.1, with Instrument Response
Function (IRF) P7\_V6, and the corresponding background files. We selected a
10$\,^{\circ}$-radius region centred on the source coordinates. \src\ and all
the sources included in the 2 LAT Catalog (\citealt{Nolan2012}) spatially
localized within this region were taken into account. The likelihood
analysis (\citealt{Mattox1996}) was applied on the LAT data both as a whole
integrated over all 5 years and by considering integration bunches
one-month long (60 bins). No detection was found in both cases, and we set
an upper limit of F($>$100 MeV)$\leq$2$\times$10$^{-9}$ ph cm$^{-2}$ s$^{-1}$ 
over the 5-year integration. This upper limit corresponds to an upper limit
of the gamma-ray monochromatic luminosity at 100~MeV of 
$\nu L_{\nu}$(100MeV)$\leq$2.1$\times$10$^{44}$ erg s$^{-1}$ which 
is within the range of gamma-ray luminosities observed
in blazars or RL NLS1 with a similar radio power (\citealt{Foschini2011}). 
Therefore it is not yet 
possible to derive any conclusion about the differences/similarities 
between \src\ and the class of ``oriented'' sources. 

\section{Discussion and conclusions}

SDSSJ143244.91+301435.3 is one of the few (13 in total 
discovered so far) radio-loud 
NLS1 with a radio-loudness parameter (R$_{1.4}$) of the order of 500 or 
greater. The radio emission, however, 
is different from that usually observed in RL NLS1 with such a high value of 
R, since it shows a steep ($\alpha$=0.93) spectrum.
Based on the size of the radio emission ($\leq$1.4 kpc)
this object is a compact radio source. 
The steep radio spectrum and the lack of strong variability and polarization, 
instead, disfavor the hypothesis that the observed compactness is due to 
the orientation of a relativistic jet towards the observer, as in blazars. 
This makes 
SDSSJ143244.91+301435.3 one of the few examples of RL NLS1 with non-blazar 
properties and sets it as an interesting case to study the 
parent population of these sources.

SDSSJ143244.91+301435.3 hosts a relatively low-mass SMBH 
(3.2$\times$10$^7$ M$_{\sun}$) accreting
at a rate close to the Eddington limit (Eddington ratio of 0.27), in 
agreement with what is 
typically found in both RQ and RL NLS1 objects.
We have also investigated the host galaxy properties through a modelling of the SED. 
We have found that the AGN dominates the emission in the visible range 
and at  2-3 $\,\mu$m while the host galaxy shows up in the 0.6-2 $\,\mu$m spectral range 
(we also detect a hint of extension in the $i$-band image) and in the MIR ($>$10 $\,\mu$m).
The best modelling is obtained using a  star-forming galaxy template (M82) with a 
SFR, estimated from the IR emission, of $\sim$50 M$_{\sun}$ y$^{-1}$. 
The presence 
and the intensity of the SF is in agreement with what is usually observed in radio-quiet NLS1 
(e.g. \citealt{Sani2010}).
Given the presence of the SF we have evaluated whether this could 
explain the observed radio emission but we have concluded that the
SF is expected to contribute only marginally ($\sim$1 per cent) to the observed 
radio fluxes. This result, combined to the large value of the radio-loudness
parameter, confirms the presence of a jet in \src.

The compact morphology and the steep radio spectrum are characteristics 
remarkably similar to those observed in the CSS radio sources.
It is nowadays clear that the large majority of CSS (and GPS)  
sources are young objects in which the radio emitting plasma is
digging its way through the ISM of the parent galaxy. Observations of
hot-spot advance speed (e.g. \citealt{Polatidis2003} and 
references therein) and radiative ages (e.g. \citealt{Murgia2003, Murgia1999}) 
for a few tens of objects show typical
ages of the order of $10^3 - 10^5$ years for GPS and CSS objects
respectively.
The association of some RL NLS1 with CSS sources has been already proposed 
by other authors (\citealt{Moran2000, Oshlack2001, Gallo2006, Komossa2006}). 
\citet{Yuan2008} also stressed the similarities between RL NLS1 and CSS sources and 
suggest that RL NLS1 with a flat radio spectrum could be CSS sources whose jet is 
pointing towards the observer.  
Another interesting characteristic that makes \src\ similar to CSS sources is the
presence of strong blue wings in the [OIII]$\lambda\lambda$4959\AA,5007\AA\ 
doublet, with FWHM of the order of 1150-1350 km s$^{-1}$ and offsets,
with respect to the core of the lines, of $\sim$250-270 km s$^{-1}$. 
The presence of these blue-shifted and  broad wings is very common in compact 
radio objects, including CSS sources (\citealt{Gelderman1994, Holt2008, 
Kim2013})
and it is usually considered as a signature of the impact of
a radio jet on the interstellar medium of the host galaxy; a jet-driven
outflow is often invoked as the best explanation for the observed
blue-shifted, broad wings of the narrow lines  (\citealt{Holt2008, 
Nesvadba2011, Kim2013}). Therefore, in sources like \src\ we might be 
witnessing the AGN-induced feedback acting in the early stages of the
evolution of a radio source.

The hypothesis of CSS sources as the parent population of RL NLS1 
would naturally explain why RL NLS1 have been detected as 
compact 
radio systems without significant extended emission (except for very 
few cases). In this framework, 
\src\ would be oriented at an intermediate angle 
between face-on and edge-on systems, the former being the flat-spectrum
blazar-like RL NLS1 and the latter appearing as CSS sources with an obscured
optical spectrum (i.e. type~2 AGN). In the first case, it is difficult
to distinguish the CSS source morphology, being the radio emission 
dominated by the beaming effects; 
in the second class of sources it is difficult, if not impossible, to
establish the NLS1 nature, due to the presence of obscuration. 
Due to the particular orientation of the source, instead, a detailed radio 
follow-up of \src\ at m.a.s. resolution can provide important pieces of information about the
intrinsic (i.e. not affected by beaming) radio properties of RL NLS1. 
A systematic search for sources like \src, 
where both NLS1 and CSS  characteristics can be observed, and a radio 
follow-up of all these objects is then mandatory to confirm 
the possible link between CSS sources and NLS1 on a firm statistical basis.

\section*{Acknowledgments}
We thank Dr.~Roberta Paladini for her kind support on the analysis of WISE data, Dr.~Ian Browne for 
useful discussions and the referee for a careful reading of the manuscript.
Part of this work is based on observations carried out at the 100-m telescope of the MPIfR 
(Max-Planck-Institut f\"ur Radioastronomie) 
at Effelsberg. This publication makes use of data products from the Sloan Digital Sky Survey (SDSS), 
from the Two Micron All Sky Survey (2mass), from the Wide-field Infrared Survey Explorer (WISE) and from
many radio surveys (NVSS, FIRST, WENSS, 7C and VLSS). 
The SDSS is managed by the 
Astrophysical Research Consortium for the Participating Institutions.
WISE is a joint project of the University of California, Los Angeles, and the Jet Propulsion 
Laboratory/California 
Institute of Technology, funded by the National Aeronautics and Space Administration. 
The 2mass is a joint project of the University of Massachusetts and the Infrared Processing and 
Analysis Center/California Institute of Technology, funded by the National Aeronautics and Space 
Administration and the National Science Foundation.
The National Radio Astronomy Observatory is a facility of the National Science Foundation operated 
under cooperative agreement by Associated Universities, Inc.
The Westerbork Synthesis Radio Telescope is operated by the Netherlands Institute for Radio 
Astronomy ASTRON, with support of NWO.

Part of this work was supported by the COST Action MP0905 ``Black Holes in a Violent Universe''.
The authors acknowledge financial support from the Italian Ministry of Education, 
Universities and Research (PRIN2010-2011, grant n.
2010NHBSBE) and from ASI (grant n. I/088/06/0). 
The research leading to these results has received funding from the European Commission Seventh 
Framework Programme (FP7/2007-2013) under grant agreement n.267251 "Astronomy Fellowships in Italy" (AstroFIt).
S. Anton acknowledges the support from FCT through project
PTDC/FIS/100170/2008 and Ciencia2007

\bibliographystyle{mn2e}
\bibliography{/home/guincho/caccia/cartella_comune/pap14323+014}
\end{document}